\documentclass[12pt, fleqn]{article}
\usepackage[cp1251]{inputenc}
\usepackage{latexsym,amsfonts,amssymb}
\usepackage{graphicx}

\usepackage{amsbsy}
\usepackage{amsmath}
\usepackage{epsf}

\newtheorem{Theorem}{Theorem}
\newtheorem{Remark}{Remark}
\newtheorem{Definition}{Definition}
\newcommand{\bt}{\begin{theo}}
\newcommand{\et}{\end{theo}}
\newcommand{\bd}{\begin{displaymath}}
\newcommand{\ed}{\end{displaymath}}

\newcommand{\lf}{\left}
\newcommand{\rg}{\right}
\newcommand{\be} {\begin{equation}}
\newcommand{\ee} {\end{equation}}
\newcommand{\ba} {\begin{array}{l}}
\newcommand{\ea} {\end{array}}
\newcommand{\bea}{\begin{eqnarray}}
\newcommand{\eea} {\end{eqnarray}}
\newcommand{\p} {\partial}
\newcommand{\vp} {\varphi}
\newcommand{\lbd} {\lambda}
\newcommand{\al} {\alpha}

\begin{document}

\begin{center}
 {\Large \bf Construction and application of  exact solutions \\ of
 the diffusive Lotka--Volterra system:\\ a review and new results }

\medskip

{\bf Roman Cherniha \footnote{\small  Corresponding author. E-mail: r.m.cherniha@gmail.com} and Vasyl' Davydovych}
 \\
{\it ~Institute of Mathematics,  National Academy
of Sciences  of Ukraine,\\
 3, Tereshchenkivs'ka Street, Kyiv 01004, Ukraine
}
 \end{center}
 \medskip
 \begin{flushright} \it  Dedicated to the memory of Wilhelm Fushchych (1936-1997) \end{flushright}


\begin{abstract} This review summarizes all  known results (up to this
date) about methods of integration of the classical Lotka--Volterra
systems with diffusion and presents a wide range  of exact
solutions, which are the most important from applicability point of
view. It is the first attempt in this direction. Because the
diffusive  Lotka--Volterra systems are used for mathematical
modeling enormous variety of processes in ecology, biology,
medicine, physics and chemistry, the review should be interesting
not only for specialists from Applied Mathematics but also those
from other branches of Science. The obtained exact solutions can
also be used as test problems for estimating the accuracy of
approximate analytical and numerical methods for solving relevant
boundary value problems.
\end{abstract}

\medskip


\section{Introduction} \label{sec-1}

About 100 years ago, Alfred Lotka \cite{lot} and Vito Volterra
\cite{vol} independently developed a mathematical model, which
nowadays serves as a mathematical background for population
dynamics, chemical reactions,
 ecology, etc. The  model is based on a system of
 ordinary differential equations
  (ODEs) involving  quadratic nonlinearities (typically two equations).
Following some earlier papers, in which {\it linear ODEs} were used
for mathematical modeling of chemical reactions (in particular, see
\cite{hirniak-1908,hirniak-1911,lotka-1910}),  Lotka has shown that
the densities
 in periodic reactions  can be adequately described by a model involving {\it ODEs with quadratic
 nonlinearities.} In contrast to Lotka,  Volterra, as a mathematician, was
 inspired by the information
  that the amount of predatory fish caught in Italy  varied  periodically
   and suggested a prey--predator model
   for the interaction of two  populations of fishes.

  The  classical Lotka--Volterra  system consists of two nonlinear ODEs of the form
\begin{equation}\label{0-1}\begin{array}{l}\frac{du}{dt}=u(a-bv),\\
\frac{dv}{dt}=v(-c+du),\end{array}\end{equation} where the functions
$u(t)$ and $v(t)$ represent
 the numbers of
prey and predators at time $t$, respectively,  $a, \ b, \ c$ and $d$
are positive parameters, the
 interpretation of which is presented below.
 Verbally,   the  Lotka and Volterra model can be formulated
  as follows:

 [change of $u$ for a small time interval] =
 [the Malthusian law (with the exponent $a$) of  growth of $u$ without predation ]
  $-$ [the quadratic  law of loss  of $u$  due to
 predation with the coefficient~$b$],

 [change of $v$ for a small time interval] = $-$ [the Malthusian law (with the exponent $-c$) of
  loss of $v$ without prey] $+$ [the quadratic  law of growth of $v$ due to
predation  with the coefficient~$b$].

Later it was shown
that  two  ODEs with quadratic nonlinearities describe  some other
types of population interaction, e.g., competition and mutualism,
 hence nowadays the   Lotka--Volterra system is
usually presented  in the form
\begin{equation}\label{0-2}\begin{array}{l} \frac{du}{dt}
=u(a_1+b_1u+c_1v),\\
 \frac{dv}{dt} =  v(a_2+b_2u+c_2v). \end{array}\end{equation}
 In particular,
  three  common types
 of interaction between two populations (predator--prey interaction,
  competition  and mutualism) can be modeled,
depending on  the signs of coefficients in (\ref{0-2}).

In the case of  interaction of $m$ species (cells, chemicals, etc.),
 a natural generalization of (\ref{0-2}) can be formulated.
Moreover, their diffusion in space  should also be  taken into
account. Thus,
 the diffusive  $m$-component Lotka--Volterra system is obtained
  \be\label{0-3}
u^i_t= d_i \Delta u^i+u^i\lf(a_i+\sum\limits_{
j=1}^{m}b_{ij}u^j\rg), \quad i=1,\dots,m, \ee where $u^1(t,x),
u^2(t,x), \ldots, u^m(t,x)$ are unknown functions, $d_i\geq0, \
a_i$ and $b_{ij}$ are arbitrary constants ($i,j=1,\dots,m),$ while
$x=(x_1,x_2,\ldots,x_n)$, $u^i_t=\frac{\p u^i}{\p t}$ and $\Delta$
is the Laplace operator  $ \frac{\partial^2}{\partial x_1^2}+\dots+
\frac{\partial^2}{\partial x_n^2} $. Nowadays the diffusive
Lotka--Volterra (DLV) system  is used as the basic model for a
variety of processes in biology, chemistry,  ecology, medicine,
economics, etc. \cite{britton, mur2, mur2003, okubo,ku-na-ei-16,
fife-79, aris-75I}. Typically, the functions  $u^j \ (j=1,\dots,m)$
are nonnegative and describe concentrations of species in
populations, cells  and drugs in tissue (tumour, bones, etc.),
chemicals in a volume.

Obviously,  this model in the case $n=1$  and $m=2$  reads as
\begin{equation}\label{0-4}\begin{array}{l}
u_t = d_1u_{xx}+u(a_1+b_1u+c_1v), \\
v_t = d_2v_{xx}+ v(a_2+b_2u+c_2v),
\end{array}\end{equation} where   the lower subscripts  $t$ and $x$ denote differentiation with respect to (w.r.t.)
these variables,  $u=u(t,x)$  and $v=v(t,x)$ are to-be-found
functions,
$a_i,\ b_i$ and $c_i$
 are given parameters (some of them can vanish and various types
 of interactions arise  depending on their
 signs),
  $d_1$ and $d_2$ are diffusion coefficients.
   If the diffusivities $d_1=d_2=0$  then (\ref{0-4})
  reduces to the classical Lotka--Volterra system.

 In the case $n=1$  and $m=3$,  the DLV system takes the form
 \begin{equation}\label{0-4*}\begin{array}{l} u_t =  d_1u_{xx}+u(a_1+b_1u+c_1v+e_1w),\\
 v_t =d_2 v_{xx}+ v(a_2+b_2u+c_2v+e_2w),\\  w_t
=d_3 w_{xx}+w(a_3+b_3u+c_3v+e_3w).
\end{array} \end{equation}
Here   $u(t,x), \ v(t,x)$  and $ \ w(t,x)$ are again  unknown
concentrations of three different  populations (cells, chemicals)
moving with diffusivities $d_1, d_2$ and $d_3$, respectively.  The
parameters  $a_k, \ b_k, \ c_k,$ and $ e_k$ define the type of
interaction between the populations. It should be noted that the
three-component models describe an essentially larger number of
interactions than those involving two components. For example, two
populations can be predators w.r.t. the third population (the
predator-predator-prey model), on the other hand, the first
population can be predators w.r.t. the other two  which  can compete
for the same food (the predator-prey-competition model).

 In contrast to the  Lotka--Volterra system  (\ref{0-2}),
 the DLV system   attracted attention of scholars much later.
 To the best of our knowledge,  its  rigorous study started
 in the 1970s
 \cite{conway,hastings,carmi,rothe}. At the present time, there
  are a lot of recent works devoted to qualitative and numerical analysis of
   the DLV system (\ref{0-4}) and  multi-component systems of the
   form  (\ref{0-3})
   (see, e.g., \cite{alha-qu-19,lam-20} and works cited therein).

However, the number of  papers devoted to the construction of exact
solutions of
 the nonlinear system (\ref{0-4}) is relatively small.
Probably the first work, in which exact solutions of the DLV system
were constructed in an explicit form, was written  by Rodrigo and
Mimura \cite{rod-mimura-2000}. The authors implicitly used a
so-called tanh-method
 \cite{malfliet-hereman-96,malfliet-04} (there are many recent papers, in which
 this method was  rediscovered without proper citations)
 for identifying  some traveling waves. In  \cite{ch-du-04}, the well-known solution of the Fisher
 equation \cite{abl-zep} was used for finding
 traveling waves of the DLV system.  Exact  solutions of the DLV system (\ref{0-4}) in the form of traveling
 waves were also constructed
 in  \cite{rod-mimura-2001,ch-du-04,hung12,kudrya-15}.
   In the case of the three-component DLV system, some traveling waves were
  found in \cite{hung11, chen-hung12,chen-hung16}, while the  existence of
traveling wave solutions  was examined in
  \cite{alha-qu-19,lam-20,hou2008,hou2011}.
  Exact solutions with more complicated structures were derived
   only in   \cite{ch-dav-2011,ch-dav-2021}  and \cite{ch-dav2013,hung16} for the two- and
    three-component DLV systems, respectively.
    In  \cite{pliukhin-15}, a natural generalization of system (\ref{0-4})
      involving additional linear and/or quadratic terms was
      studied and its  exact solutions were derived.
     It should  also be mentioned that systems of nonlinear ODEs for constructing exact solutions of (\ref{0-4}) in the very special case
     when $a_1=a_2,\ b_1=b_2, \ c_1=c_2, \ d_1=d_2$  are presented
     in the handbook \cite{polyanin-2012}.
     However,  relevant exact solutions are not presented therein.

Hence, the problem of construction of exact solutions of   DLV
systems, especially those with a biological,   chemical or physical
interpretation, is a hot topic.
 Notably construction of exact solutions with more complicated  structures
 (compared to
 traveling waves)
  requires more sophisticated  methods and techniques.

 At the present time, the most useful  methods for construction of exact solutions for
nonintegrable nonlinear partial differential equations (PDEs) are
symmetry-based methods (see Section~\ref{sec-2}). These methods are
based on
 the Lie method, which was created by a famous Norwegian
mathematician Sophus Lie in the late  19th century.  The Lie method
(the notions `the Lie symmetry analysis', `the Lie group analysis'
and `the group-theoretical analysis' are  used  as well) still
attracts  attention of many investigators and new results are
published on a regular basis (see the recent monographs
\cite{bl-anco-10, ch-se-pl-book} and papers cited therein). On the
other hand, many  nonlinear PDEs and systems of PDEs arising in
real-world  applications have a poor Lie symmetry. The Lie method is
not productive for such type equations  since in this case exact
solutions can be easily obtained without using this cumbersome
method.
be easily obtained without using this cumbersome  method.
 The DLV system (\ref{0-4}) is a  typical example of such type systems
because one admits a nontrivial Lie symmetry
 only under essential  restrictions on the coefficients
 $a_i, \ b_i$ and $c_i$
  (see   Section~\ref{sec-3}).
As a result,  direct application of the Lie method  leads  only
traveling wave solutions for (\ref{0-4}) (see Section~\ref{sec-4}),
otherwise some coefficients in (\ref{0-4}) must vanish.

 Within recent  decades, new
symmetry-based methods were developed  in order to solve nonlinear
PDEs arising in applications, but possessing  poor Lie symmetry.
   The
 method  of nonclassical  symmetries proposed by Bluman and Cole in
 1969 \cite{bl-co-69} is one of
 the best-known among   them. It should be noted that we
  use the terminology   `$Q$-conditional symmetry'
  instead of `nonclassical symmetry'. In our opinion,
  this terminology,  proposed by  Fushchych  in the 1980s \cite{fu-se-ch-88,
  fss} when he and his collaborators proposed a generalization of nonclassical
  symmetries, more adequately reflects the essence of the  method.

   Although the method suggested in \cite{bl-co-69} is rather simple,
   its  successful applications for
  solving {\it nonlinear  systems
   of PDEs}  were accomplished only in the 2000s.
   Moreover, the majority of the papers devoted to conditional symmetries of
   reaction-diffusion systems (the DLV system (\ref{0-4}) is a typical example)
    were published  within the recent decade
  \cite{ch-dav-2011,ch-dav2013,arrigo2010,ch-2010,torissi,che-dav2015,ch-dav2021}.
  It happened so late because application
  of the nonclassical method \cite{bl-co-69} to nonlinear  systems
   of PDEs leads
  to very complicated nonlinear overdetermined systems of  differential  equations
   to-be-solved. In other words, one needs to solve  a much  more complicated
   PDE system (a so-called system of determining equations (DEs)) comparing
   with   the initial  system of PDEs.
    In   order to make essential  progress in solving systems
    of DEs, a new definitions of $Q$-conditional symmetries  and
    new algorithm were proposed in  \cite{ch-2010}.
     The algorithm follows from the  notion of a {\it $Q$-conditional symmetry
    of the first type}. In recent papers, we successfully applied this algorithm
    for constructing new exact
     solutions of the DLV system (\ref{0-4}) and (\ref{0-4*}) (see
     Sections~\ref{sec-5}, \ref{sec-6} and~\ref{sec-7}).
     In Sections~\ref{sec-4}, \ref{sec-6} and~\ref{sec-7}, we also
     discuss  the most interesting exact solutions derived by other
     authors using  other techniques.


\section{Main definitions} \label{sec-2}

Let us consider the DLV system (\ref{0-3}).  First of all, we note  that the DLV system (\ref{0-3}) with $d_i>0 \ (i=1,\ldots,m)$ in the $(1+1)$-dimensional case can be rewritten as
 \be\label{1-2}
\lambda_iu^i_t= u^i_{xx}+u^i\lf(a_i+\sum\limits_{
j=1}^{m}b_{ij}u^j\rg), \quad i=1,\dots,m, \ee by introducing the
notations $d_i \rightarrow \frac{1}{\lambda_i}, \ a_i \rightarrow
\frac{a_i}{\lambda_i}, \ b_{ij} \rightarrow
\frac{b_{ij}}{\lambda_i}.$ In what follows we study the DLV system
in the form (\ref{1-2}) because the terms
 with the higher-order derivatives do not involve any coefficients.
  Obviously, each result obtained for system (\ref{1-2}) is valid  for the DLV
  system (\ref{0-3}) after introduction of the inverse notations
  \[\lambda_i \rightarrow \frac{1}{d_i}, \ a_i \rightarrow
\frac{a_i}{d_i}, \ b_{ij} \rightarrow \frac{b_{ij}}{d_i}.\]

Plane wave solutions form the  most common  class  of the exact solutions  of two-dimensional PDEs (system of PDEs)
  because such solutions are important from an applicability point of view. In particular, traveling fronts,
   i.e. plane wave solutions, which are nonnegative, bounded and satisfy the zero Neumann conditions at infinity,
    are the most interesting solutions for a wide range of applications.
 Properties of such  solutions in the case of   scalar nonlinear  reaction-diffusion (RD) equations
 were extensively studied during the recent decades using different mathematical techniques
 (see, e.g., monographs \cite{ch-se-pl-book,gi-ke-04} and references cited therein).In the case of systems
 of RD equations, the progress is rather modest  especially  in searching for the plane wave solutions
 in explicit forms (the main references are listed in Introduction).
 The corresponding   ansatz (a special kind of  substitutions)  for search for plane waves of  a given $m$-component
  RD system (including the DLV system)  has the form
 \be\label{1-2*}
u^i=\vp_i(\omega), \ i=1,\dots,m, \,  \omega=x- \mu t, \ee where
$\vp_i$ are to-be-determined functions and  the parameter $\mu$
means the wave speed. Obviously, each system of (1+1)-dimensional RD
equations with coefficients that do not depend explicitly on time
$t$ and space $x$, is reducible to a system of ODEs via ansatz
(\ref{1-2*}). The system obtained does not depend on the new
variable $\omega$. There are many techniques for solving  such
systems of ODEs, however, their applicability depends essentially on
the structure of the system in question.  We will demonstrate this
in Section~\ref{sec-4} for the ODE systems corresponding to the DLV
system (\ref{1-2}).

The symmetry-based methods allow us to construct ans\"atze with more
complicated structures than (\ref{1-2*}). It turns out that each new
ansatz obtained  via a symmetry also   reduces the given RD system
to a system of ODEs although  the relevant reduction can be highly
nontrivial in contrast to the reduction via (\ref{1-2*}). In what
follows we restrict ourselves by the classical Lie method and the
method of $Q$-conditional symmetries \cite{ch-se-pl-book,fss}. Both
methods allow us to construct ans\"atze of the form
 \be\label{1-2**}
  u^i = g_i(t,x)+ G_{ij}(t,x)\vp_j\bigr(\omega(t,x)\bigr), \ i=1,\dots,m, \ee
  provided  the $m$-component  RD system in question admits a Lie and/or $Q$-conditional symmetry.
Here $\vp_i$ are to-be-determined functions of   the variable
$\omega(t,x)$, while  $g_i(t,x)$  and   $G_{ij}(t,x)$ are the known
functions and a summation is assumed from 1 to $m$ over the repeated
index $j$. Obviously, formulae (\ref{1-2*}) follow from
(\ref{1-2**}) as a very particular case.

Thus, in order to derive new reductions of   the DLV system
(\ref{1-2}), one needs to construct its Lie and $Q$-conditional
symmetries enabling    ans\"atze  of the  form(\ref{1-2**}) to be
found. Any Lie and $Q$-conditional symmetry  has the form of  a
linear first-order differential operator (infinitesimal operator)
 \be\label{2-1}\ba
Q = \xi^0 (t, x, u^1,\ldots,u^m)\p_{t} + \xi^1 (t, x,u^1,\ldots,u^m)\p_{x} + \\
 \hskip3cm \eta^1(t, x,u^1,\ldots,u^m)\p_{u^1}+\ldots+\eta^m(t, x,u^1,\ldots,u^m)\p_{u^m}  \ea \ee
 with the correctly specified coefficients $\xi^0, \ \xi^1, \
 \eta^1,\ldots,\eta^m$. Hereinafter we use the notations  $\p_z=\frac{\p
}{\p z},  \  z=t,x,u^i,...$ .

It is well-known
 that in order  to find a Lie symmetry of system (\ref{1-2}), one needs to
consider  the system as the manifold
\[{\cal{M}}=\lf\{S_1=0,S_2=0, \dots, S_m=0 \rg\},\] where \[
 S_i \equiv \lambda_iu^i_t-u^i_{xx}-u^i\lf(a_i+\sum\limits_{
j=1}^{m}b_{ij}u^j\rg), \ i =
1,2, \ldots, m, \] in the prolonged space of the variables
\[t,\ x,\ u^i, \ u^i_t, \ u^i_x, \ u^i_{tt}, \ u^i_{tx},
\ u^i_{xx}, \ i =
1,2, \ldots, m. \]

\begin{Definition}\label{d0} The infinitesimal
operator  (\ref{2-1})  is a Lie symmetry of  system (\ref{1-2}) (in other words the latter
 is invariant  under the transformations generated by  (\ref{2-1}))  if the following invariance
  criterion  is satisfied:
\be\label{2-2}  
\mbox{\raisebox{-1.6ex}{$\stackrel{\displaystyle  
Q}{\scriptstyle 2}$}}\, (S_i)\Big\vert_{\cal{M}}=0,  \ i = 1,2,
\ldots, m.
\ee  
\end{Definition}

The operator $ \mbox{\raisebox{-1.6ex}{$\stackrel{\displaystyle  
Q}{\scriptstyle 2}$}}$ is the second-order prolongation of the
operator $Q$ and its coefficients are expressed via the functions
$\xi^0, \ \xi^1, \ \eta^1,\ldots,\eta^m$ by the well-known formulae
(see  any texbook/monograph  devoted to  Lie symmetries of PDEs).

The main idea  used for introducing the notion of the
$Q$-conditional symmetry
  is to change  the manifold
${\cal{M}}$ in formulae  (\ref{2-2}).
 It was  noted in  \cite{ch-2010} that there
are several different possibilities to modify the manifold
${\cal{M}}$ in the case of PDE systems. The first possibility is
natural and follows directly from the seminal work \cite{bl-co-69}
(see more details in \cite{bl-anco}).

\begin{Definition}\label{d1}  Operator  (\ref{2-1}) is called a
$Q$-conditional symmetry (nonclassical symmetry)
for DLV system (\ref{1-2}) if the following
invariance criterion  is  satisfied:
\be\label{2-3}
\mbox{\raisebox{-1.6ex}{$\stackrel{\displaystyle  
Q}{\scriptstyle 2}$}}\, (S_i)\Big\vert_{{\cal{M}}_m}  
  =0,  \ i = 1,2,
\ldots, m. \ee Here  the manifold has the form
  \[{\cal{M}}_m = \lf\{S_i=0,Q\left(u^{i}\right)=0,  \frac{\partial}{\partial t}\,Q\left(u^{i}\right)=0, \frac{\partial}{\partial x}\,Q\left(u^{i}\right)=0,i=1,\dots,m\rg\},\] where \[Q\left(u^{i}\right)\equiv \xi^0{u^i}_t+\xi^1{u^i}_x-\eta^i.\]
\end{Definition}

Another possibility is to consider a manifold  ${\cal{M}}_*$ , which is between
${\cal{M}}$  and ${\cal{M}}_m$, i.e.
\[ {\cal{M}} \supset {\cal{M}}_*\supset {\cal{M}}_m.  \]
There are several possibilities and the simplest one is
 \[ {\cal{M}}_*={\cal{M}}^j_1=\lf\{S_1=0,S_2=0, \dots, S_m=0, Q\left(u^{j}\right)=0,  \frac{\partial}{\partial t}\,Q\left(u^{j}\right)=0, \frac{\partial}{\partial x}\,Q\left(u^{j}\right)=0
\rg\},\]
where $j  \ ( 1\leq j \leq m)$ is {\it a fixed number}.

\begin{Definition}\label{d2}
Operator  (\ref{2-1}) is called $Q$-conditional symmetry
of the first type for DLV system (\ref{1-2}) if the following invariance criterion
 is  satisfied:

\be\label{2-4}
\mbox{\raisebox{-1.6ex}{$\stackrel{\displaystyle  
Q}{\scriptstyle 2}$}}\, (S_i)\Big\vert_{{\cal{M}}^j_1}=0,  \ i = 1,2,
\ldots, m. \ee
\end{Definition}

In the case of the DLV system (\ref{1-2}) with $m>2$,  there are more possibilities to construct new manifolds  ${\cal{M}}_*$ (see  \cite{ch-2010,ch-dav-book} for details).

The algorithms for search  Lie and $Q$-conditional symmetries  of the DLV system (\ref{1-2})  are based on the definitions presented above  and the standard methods for solving overdetermined systems of PDEs  and linear systems of differential equations.

\section{\bf Lie symmetries of the DLV systems}\label{sec-3}

The prominent Norwegian  mathematician Sophus Lie was the first to develop and apply the method for finding Lie
 symmetries of PDEs. Nowadays this method is well-known and can
 be found together with examples in many monographs and textbooks
 (the most recent are
 \cite{bl-anco-10, ch-se-pl-book,bl-anco,ch-dav-book,1-arrigo15}).
Here we present  results of its  application  to the DLV systems
skipping excessive  details. 

 System (\ref{1-2}) in the case
$m=2$, i.e. a two-component DLV system, has the form (up to the
notations)
\begin{equation}\label{3-1}\begin{array}{l} \lbd_1 u_t = u_{xx}+u(a_1+b_1u+c_1v),\\ \lbd_2 v_t =
v_{xx}+ v(a_2+b_2u+c_2v). \end{array}\end{equation}

Here we examine system (\ref{3-1}) assuming that both equations
 are nonlinear and  not autonomous, i.e.
 \be\label{3-6} b_1^2+c_1^2\neq0,  \ b_2^2+c_2^2\neq0,  \ c_1^2+b_2^2\neq0.  \ee
 The above restrictions are natural. In fact,  assuming
 $c_1^2+b_2^2=0$, one obtains two autonomous equations, which cannot
 describe any kind of interaction between species (cells,
 chemicals). Setting $b_1^2+c_1^2=0$  (or $b_2^2+c_2^2=0$), we arrive
 at a system involving the linear diffusion equation with a
 linear source/sink. Such a system  is not interesting from both the
 mathematical  and applicability
 point of view.

It is obvious that   the DLV system (\ref{3-1}) with arbitrary
coefficients admits a  two-dimensional Lie algebra generated by the
operators
 \be\label{3-2}P_t=\p_t, \quad P_x=\p_x.\ee   Obviously,
 the above operators generate the following invariance
transformations of (\ref{3-1}):
\[  t^*=t+t_0,  \quad   x^*=x+x_0,\]
where $t_0$  and $x_0$  are arbitrary parameters.


It turns out  that there are several cases when this nonlinear
system with correctly-specified coefficients is invariant w.r.t. a
three- and higher-dimensional Lie algebra.

\begin{Theorem}\label{3-t1}\cite{ch-du-04}  The DLV system (\ref{3-1}) with
restrictions (\ref{3-6}) admits three- and higher-dimensional
 Lie algebra if and only if its nonlinear terms  and the
corresponding  symmetry operator(s) have  structures listed in
Table~\ref{3-tab1}. If the DLV system (\ref{3-1}) with other
reaction terms is invariant w.r.t. a nontrivial Lie algebra, then it
is reduced to one of the forms presented in Table~\ref{3-tab1} by a
 substitution of the form
 \[
 u \to c_{11} \exp(c_{10} t) u+c_{12}, \quad v \to
c_{21}+ c_{22} \exp(c_{20} t) v\] (here $c_{ki} $ ($k=1,2,\
i=0,1,2,$) are  correctly-specified parameters).
 \end{Theorem}

\begin{table}[h!]
\caption{ Lie symmetries of the DLV system (\ref{3-1}).} \label{3-tab1}
\begin{center}
\begin{tabular}{|c|c|c|c|} \hline &&&\\
 &  Reaction terms & Restriction &  Lie
 symmetries extending
  algebra (\ref{3-2}) \\
\hline
 &&& \\
1 & $u( b_1u +  c_1u )$ & & $D = 2tP_t + x P_x -2(u \p_u + v \p_v)$\\
&$v( b_2u + c_2v)$&& \\ \hline &&& \\
2 & $b_1u^2$ & & $D, \ v\p_v$ \\
&$b_2uv$&&  \\ \hline &&& \\
3 & $u(a_1+b_1u)$ & & $v\p_v$\\
&$b_2uv$&&  \\ \hline &&& \\
4 & $u(a_1+ b_1u)$ & $\lbd_1=\lbd_2$ & $v\p_v, \ u\p_v, \ (a_1+b_1u)e^{a_1t}\p_v$\\
&$v(a_1+ b_1u)$&&  \\ \hline &&& \\
5 & $b_1u^2$  & $\lbd_1=\lbd_2$ & $v\p_v, \ u\p_v, \ D, \ R= b_1tu \p_u+\p_v$\\
&$b_1uv$&&  \\ \hline
  \end{tabular}
\end{center}
\end{table}

It can be seen from Table~\ref{3-tab1} that the  DLV systems
possessing nontrivial Lie symmetry  are semi-coupled (see Cases
2--5), except for Case 1. From the applicability point of view, the
DLV system (\ref{3-1}) with $a_1=a_2=0$  is the most important among
others.

System (\ref{1-2}) in the case $m=3$, i.e., three-component DLV
system, has the form (up to the notations)
 \be\label{3-3}\ba \lambda_1 u_t =  u_{xx}+u(a_1+b_1u+c_1v+e_1w),\\
\lambda_2 v_t = v_{xx}+ v(a_2+b_2u+c_2v+e_2w),\\ \lambda_3 w_t
= w_{xx}+w(a_3+b_3u+c_3v+e_3w). \ea\ee

Note that we want to exclude   the system  containing an autonomous
equation from the study, hence, hereinafter  the restrictions
\begin{equation}\label{3-4} c^2_1+e^2_1\neq0, \  b^2_2+e^2_2\neq0, \
b^2_3+c^2_3\neq0\end{equation} are assumed. Similarly to the
two-component case,  the above restrictions are natural from the
applicability point of view.

 A complete description of  Lie symmetries of the three-component DLV system  (\ref{3-3}) was derived in \cite{ch-dav2013}.
  Obviously, the DLV system (\ref{3-3}) with arbitrary coefficients $a_k, \ b_k, \ c_k, \
e_k$ and $\lambda_k$ admits the Lie algebra with the basic operators
(\ref{3-2}).
In order to find all possible extensions of the Lie algebra
(\ref{3-2}),
 it is necessary to apply the invariance criterion (\ref{2-2}),
 to solve the DEs obtained    and to identify  all
possible   restrictions on the  coefficients $\lambda_i, ..., e_i$
leading to extensions of the Lie algebra (\ref{3-2}). Because all
the coefficients of the DLV system (\ref{3-3}) are constants, this
problem can be solved by standard calculations (see section 3.3 in
\cite{ch-dav2013} for details).

\begin{table}[h!]
\caption{Lie symmetries of the DLV system (\ref{3-3})}\label{3-tab2}
\begin{center}
\begin{tabular}{|c|c|c|c|}  \hline &&&\\
& Reaction terms &Restrictions & Lie symmetries extending algebra (\ref{3-2}) \\ \hline &&&\\
 1 &
$u(b_1u+c_1v+e_1w)$  & & $D=2t\p_t+x\p_x-2(u\p_u+v\p_v+w\p_w)$ \\
&$v(b_2u+c_2v+e_2w)$&& \\
&$w(b_3u+c_3v+e_3w)$&& \\
  \hline &&&\\
2 & $u(c_1v+e_1w)$  &  & $u\p_u$\\
&$v(a_2+c_2v+w)$&& \\
&$w(a_3+v+e_3w)$&& \\
  \hline &&&\\
3 & $u(c_1v+e_1w)$ &  & $u\p_u, \ D$\\
&$v(c_2v+w)$&& \\
&$w(v+e_3w)$&& \\
  \hline &&&\\
4 &$u(a_1+bu+v)$ &
$\lambda_2=\lambda_3=1$ &
$\exp(-a_2t)v\p_w, \ w\p_w$\\
&$v(a_2+u+cv)$&& \\
&$w(u+cv)$&& \\
  \hline &&&\\
5 & $u(bu+v)$ & $\lambda_2=\lambda_3=1$ & $v\p_w, \ w\p_w, \ D$\\
&$v(u+cv)$&& \\
&$w(u+cv)$&& \\
  \hline &&&\\
6 & $u(a_1+u+v)$  &
$\lambda_1=\lambda_2=\lambda_3=1,$ &
  $\exp(-a_1t)u\p_w,$ $w\p_w, \ \exp(-a_2t)v\p_w,$ \\
  &$v(a_2+u+v)$&$a_1a_2(a_1-a_2)\neq0$& $(a_2(u+a_1)+a_1v)\p_w$\\
&$w(u+v)$&& \\
  \hline &&&\\
7 &$u(a+u+v)$  &
$\lambda_1=\lambda_2=\lambda_3=1,$&
$\exp(-at)u\p_w,$
$w\p_w, \ v\p_w, \ (u+a+avt)\p_w$\\
&$v(u+v)$&$a\neq0$& \\
&$w(u+v)$&& \\
  \hline &&&\\
8 &   $u(bu+v)$  &
$\lambda_1=\lambda_2=\lambda_3=1,$
&
 $w\p_w, \ \lf((b-1)u+(1-c)v\rg)\p_w, \ D$ \\
 &$v(u+cv)$&$(b-1)^2+(c-1)^2\neq0 $& \\
&$w(bu+cv)$&& \\
  \hline
\end{tabular}
\end{center}
\end{table}

   \begin{Theorem} \label{3-t2} \cite{ch-dav2013} The  DLV system (\ref{3-3}) with restrictions (\ref{3-4})
  admits a nontrivial Lie algebra of  symmetries if and only if
    the system  and the
corresponding Lie  symmetry operators have the forms listed in
Table~\ref{3-tab2}. Any other DLV system admitting three- and
higher-dimensional Lie algebra  is reducible to one of those from
Table~\ref{3-tab2} by a transformation from the set:

\be\label{3-5}\begin{array}{l} u \rightarrow
c_{11}\exp(c_{10}t)u+c_{12}v+c_{13}w,\\ v \rightarrow
c_{21}\exp(c_{20}t)v+c_{22}u+c_{23}w,\\ w \rightarrow
c_{31}\exp(c_{30}t)w+c_{32}u+c_{33}v, \\ t \rightarrow
c_{40}t+c_{41}, \ x \rightarrow c_{50}x+c_{51},
\end{array}\ee where $c_{ij}$ ($i=1,\dots,5$,
$j=0,\dots,3$) are correctly-specified constants (some of them
vanish) that  are defined by the DLV system in question.
\end{Theorem}

It can be seen from Table~\ref{3-tab2} that the  DLV system
(\ref{3-3}) admits  three- and higher-order Lie algebra provided at
least three coefficients vanish. It is questionable  that such
systems can arise in real-world applications. On the other hand,
some of them
result from an approximation of relevant models, e.g., the DLV
system with $a_1=a_2=a_3=0$  (see Case 1) assumes   zero natural
birth/death rate for interacting  species. It means an assumption
on the equality of natural death rate and birth rate for each
species.

\section{\bf Traveling wave solutions of the DLV systems}\label{sec-4}

In this section, we look for traveling wave solutions of the DLV
systems (\ref{3-1}) and (\ref{3-3}). Because the DLV systems
(\ref{3-1}) and (\ref{3-3}) with arbitrary coefficients admits only
the trivial algebra (\ref{3-2}), the plane wave ansatz (\ref{1-2*})
can be easily derived. In fact, if one takes a linear combination of
the operators  (\ref{3-2}) $Q=P_t +\mu P_x$ ($\mu$ is a wave speed)
and constructs the invariance surface condition
\[ Q(u^i)\equiv u_t^i +\mu u^i_x =0 \]
then ansatz (\ref{1-2*})  is immediately obtained.

Ansatz (\ref{1-2*}) with $m=2$ and $m=3$ reduces the DLV systems
(\ref{3-1}) and (\ref{3-3}) to the nonlinear ODE systems
\be\label{4-6}
\begin{array}{l}
 \varphi_1''+\alpha\lambda_1
\varphi_1'+\varphi_1(a_1+b_1
\varphi_1+c_1\varphi_2)=0, \\
 \varphi_2''+ \alpha\lambda_2\varphi_2'+\varphi_2(a_2+b_2
\varphi_1+c_2\varphi_2)=0
\end{array}
\ee
and
\be\label{4-21}
\begin{array}{l}
 \varphi_1''+\alpha\lambda_1
\varphi_1'+\varphi_1(a_1+b_1
\varphi_1+c_1\varphi_2+e_1\varphi_3)=0, \\
 \varphi_2''+ \alpha\lambda_2\varphi_2'+\varphi_2(a_2+b_2
\varphi_1+c_2\varphi_2+e_2\varphi_3)=0,\\
\varphi_3''+ \alpha\lambda_3\varphi_3'+\varphi_3(a_3+b_3
\varphi_1+c_3\varphi_2+e_3\varphi_3)=0
\end{array}
\ee (hereinafter the upper sign  $'$ denotes the derivation
$\frac{d}{d \omega}$), respectively.

To the best of our knowledge,  the ODE systems (\ref{4-6}) and
(\ref{4-21}) with arbitrary coefficients are not integrable. As a
result,  the recently published handbooks devoted to nonlinear ODEs,
e.g., \cite {pol-za}, do not contain their general solutions. Their
exact solutions (solutions in closed forms) can be derived only
under additional restrictions on parameters. For long time, these
systems were  studied using only qualitative and numerical methods.
The papers devoted to search for exact solutions, especially those
leading to traveling waves, were published only within the recent
two decades.

A majority of the papers
\cite{rod-mimura-2000,ch-du-04,hung12,kudrya-15} devoted to search
for the traveling wave solutions of the DLV system (\ref{3-1}) are
focused  on the case when the system describes  competition between
two populations (cells, chemicals). It means that the signs of the
parameters are fixed. Thus, introducing the new notations $b_k
\rightarrow -b_k$, $c_k \rightarrow -c_k$ we rewrite the DLV system
(\ref{3-1}) in the form
\be \label{4-9}
\begin{array}{l}
\lambda_1u_t=u_{xx}+u(a_1-b_1u-c_1v), \\
\lambda_2v_t=v_{xx}+v(a_2-b_2u-c_2v),
\end{array}\ee where  the coefficients $a_i, \ b_i$ and $c_i$ are nonnegative.
 The reduced ODE system corresponding to  the DLV system (\ref{4-9})  takes the
form
 \be\label{4-10}
\begin{array}{l}
 \varphi_1''+\alpha\lambda_1
\varphi_1'+\varphi_1(a_1-b_1
\varphi_1-c_1\varphi_2)=0, \\
 \varphi_2''+ \alpha\lambda_2\varphi_2'+\varphi_2(a_2-b_2
\varphi_1-c_2\varphi_2)=0.
\end{array}
\ee

As was mentioned above,  \cite{rod-mimura-2000} is the first study,
in which exact solutions of  the two-component DLV system were
constructed. In order to solve the ODE system (\ref{4-10}), the
nonlocal ansatz \cite{rod-mimura-2000}
\[\varphi_1'=\sum\limits_{i=0}^{m}\alpha_i\varphi_1^i, \ \varphi_2'=\sum\limits_{i=0}^{n}\beta_i\varphi_1^i, \ m, \ n>0\]
was used. Actually, after substitution of the ansatz into
(\ref{4-10}), the authors studied the special cases $m=1,2$ and
$n=1,2$, which naturally lead to  solutions in the form of
tanh-functions (or coth-function). So, the authors used the
tanh-method, which was developed earlier for  similar purposes
\cite{malfliet-hereman-96,malfliet-04}. Here we present the main
 exact solutions obtained in \cite{rod-mimura-2000}.

Traveling wave solutions of the DLV system (\ref{4-9}) with the
parameters
\[\lambda_1=1, \ \lambda_2=\lambda, \ a_1=1, \ a_2=a, \ b_1=1, \ b_2=2\lambda+\frac{5a}{3}-\frac{a\lambda}{3}, \ c_1=\frac{1}{3}, \ c_2=1\]  and
\[\lambda_1=1, \ \lambda_2=\frac{1+a(c-6)}{5-ac}, \ a_1=1, \ a_2=a, \ b_1=1, \ b_2=ac+1-a, \ c_1=c, \ c_2=1\]
given by  \cite{rod-mimura-2000}
\be\label{4-1}\ba u(t,x)=\frac{1}{2}\Big[1+\tanh\Big(\frac{\sqrt{a}}{2\sqrt{6}}\left(x-\frac{a-6}{\sqrt{6a}}\,t\right)\Big)\Big],\medskip\\
v(t,x)=\frac{a}{4}\Big[1-\tanh\Big(\frac{\sqrt{a}}{2\sqrt{6}}\left(x-\frac{a-6}{\sqrt{6a}}\,t\right)\Big)\Big]^2,\ea\ee
and
\be\label{4-2}\ba u(t,x)=\frac{1}{4}\Big[1+\tanh\Big(\frac{\sqrt{1+ac}}{2\sqrt{6}}\left(x-\frac{ac-5}{\sqrt{6+6ac}}\,t\right)\Big)\Big]^2,\medskip\\
v(t,x)=\frac{a}{4}\Big[1-\tanh\Big(\frac{\sqrt{1+ac}}{2\sqrt{6}}\left(x-\frac{ac-5}{\sqrt{6+6ac}}\,t\right)\Big)\Big]^2,\ea\ee
respectively.

Solution (\ref{4-1}) and (\ref{4-2}) are  typical  traveling fronts,
which are positive and bounded for arbitrary $x$ and $t\geq0$.

In  \cite{ch-du-04},  exact solutions of the ODE system (\ref{4-10})
were constructed using
 the following condition:
 \be \label{4-7} \varphi_2=\beta_0+\beta_1 \varphi_1, \ee where
$\beta_0$ and $\beta_1$ are to-be-determined  constants.
Substituting (\ref{4-7}) into (\ref{4-10}), one obtains an
overdetermined system, which possesses nonconstant solutions only
under the restriction $\lambda_1=\lambda_2=\lambda$. Without loss of
generality one can set $\lambda=1$.   Thus, the  second-order ODE
 \be
\label{4-8}
 \varphi_1'' + \alpha \varphi_1'+\varphi_1(a-b\varphi_1)=0
\ee is obtained. Here the constants $a$ and $b$ depend  on an
additional parameter $\beta_0$  as follows \be \label{4-11} a=
\left\{
\begin{array}{l}
a_1=a_2, \quad \beta_0=0,\\
a_1-a_2 \frac{c_1}{c_2}, \quad \beta_0=\frac{a_2}{c_2},
\end{array} \right. \quad
 b= \left\{\begin{array}{l}
\frac{c_1b_2-b_1c_2}{c_1-c_2}, \quad \beta_0=0,\\
b_1+c_1 \beta_1, \quad \beta_0=\frac{a_2}{c_2},
\end{array} \right.
\ee \be \label{4-12} \beta_1= \left\{ \begin{array}{l}
\frac{b_1-b_2}{c_2-c_1}, \quad c_1\not=c_2, \quad b_1 \not=b_2, \\
-\frac{a_2 b_1}{a_1 c_1}, \quad c_1=c_2, \quad b_1=b_2.
\end{array} \right.
\ee

ODE (\ref{4-8}) is known as  the reduced equation of the
famous  Fisher equation~\cite{fi-37}. In particular, ODE (\ref{4-8})
has the exact solution \cite{abl-zep} \be \label{4-13}
\varphi_1=\frac{a}{b}\left(1 + c \exp\left(\pm \sqrt{\frac{a}{6}}
\omega\right)\right)^{-2}, \ee where $\alpha=
\frac{5\sqrt{a}}{\sqrt{6}}$ and $c$ is an arbitrary constant.

 Assuming  $c>0$, taking into account formulae
 (\ref{4-7})  and (\ref{1-2*}) and  fixing  the upper sign in
  (\ref{4-13}),  one obtains the exact  solution in the form of traveling front
  \be \label{4-14}
\begin{array}{l}
u=\frac{a}{4b}\left(1-\tanh \left(\sqrt{\frac{a}{24}}
x-\frac{5a}{12}t\right)\right)^2,\\ \hskip1.37cm v=\beta_0+\beta_1
u.
\end{array}
\ee    Here the parameters $a$, $b$, $\beta_0$ and $\beta_1$ are
defined by (\ref{4-11}) and (\ref{4-12}).

We want to point out that the traveling wave solution  (\ref{4-14})
was  much later rediscovered in \cite{kudrya-15} (see formulae (18)
and (24) therein).
 Notably this solution  has essentially different properties depending on the value of the parameter $\beta_0$, therefore one
simulates different types of interaction between population (see
examples  below).

Setting   $c<0$ we observe that the exact  solution    (\ref{4-13})
generates the following solution of the DLV system (\ref{4-9}):
 \[\ba
u=\frac{a}{4b}\left(1-\coth \left(\sqrt{\frac{a}{24}}
x-\frac{5a}{12}t\right)\right)^2,\\v=\beta_0+\beta_1 u. \ea\] In
contrast to (\ref{4-14}), this solution  blows up at all points
$(t,x)$  belonging to  the plane
\[\sqrt{\frac{a}{24}}\,x-\frac{5a}{12}\,t=0.\] Probably, solutions of such
type  may describe  an unusual  interaction  when both populations
grow unboundedly.

Interestingly, that  \cite{3-gudkov} is devoted to  a special case
of the DLV system (\ref{4-9})  with $a_2=c_2=0$, i.e. a so-called
   Belousov--Zhabotinskii system.
 The  exact solution
constructed in \cite{3-gudkov} can  be obtained from (\ref{4-14})
(for details see  \cite{ch-du-04}).

An important feature
of traveling waves follows from their
property to satisfy no-flux conditions at infinity. No-flux
conditions at boundaries are typical requirements for a wide range
of real-world processes. As an example,  we use the exact   solution
(\ref{4-14}) for solving
 the Neumann boundary value problem (BVP)
 for the  DLV system
(\ref{4-9}).

\begin{Theorem} \label{4-t1} \cite{ch-du-04}
 Let us consider  the Neumann  BVP  with the governing  equations
 (\ref{4-9}),
    the initial conditions
\be\label{4-15}
\begin{array}{l}
u=\frac{a}{4b}\left(1-\tanh \left(\sqrt{\frac{a}{24}}
x\right)\right)^2 \equiv u_0(x),\\ \hskip1.37cm v=\beta_0+\beta_1
u_0(x)
\end{array} \ee and the  Neumann conditions  at infinity \be\label{4-16}
 u_x(t, -\infty)=u_x(t, +\infty)= v_x(t, -\infty)=v_x(t, +\infty)=0   \ee
in the domain $\Omega=\left\{ (t,x) \in (0,+ \infty )\times
(-\infty,+ \infty)\rg\} $ Then its  bounded    solution has the form
(\ref{4-14}).

In formulae  (\ref{4-15}) and (\ref{4-14}), the coefficients $a$,
$b$, $\beta_0$ and $\beta_1$ are defined by (\ref{4-11}) and
(\ref{4-12}).
\end{Theorem}

Now we want to suggest an example of   biological interpretation of
this theorem. First of all, we observe   that
 two essentially different cases occur, namely: $\beta_0 \not =0 $
and $ \beta_0=0$.   If $\beta_0\not=0$   then solution (\ref{4-14})
 has  the  asymptotical behavior
\be\label{4-17} (u,\,v)  \rightarrow  \biggl({a_1 \over
b_1},\,0\biggr) \quad \mbox{as} \quad  t  \rightarrow \infty, \ee
 provided  the following  condition is satisfied:
 \be\label{4-18} A>
\max\{B,C\}, \ee where $A={a_1 \over a_2}, \ B={b_1 \over b_2}, \
C={c_1 \over c_2}$ (note that the condition $A(B-1)=B(C-1)$ follows
from  (\ref{4-14}) and (\ref{4-17})). In population dynamics, such
asymptotical behavior  predicts  an uncompromising competition
between two populations of species $u$ and $v$.  In other words, any
increase in population
 $u$ leads to a decrease in species
$v$. As a result, the  species $v$ completely disappear.

It turns out that  the opposite condition
 \be\label{4-19} A<
\min\{B,C\} \ee  leads to  the competition with the same character.
In this case, the species $v$ dominates, while the species $u$
eventually dies out.

If $\beta_0=0$ (in this case, the restriction  $a_1=a_2=a$ follows
from (\ref{4-11}))  then solution (\ref{4-14}) possesses the
property
 \be\label{4-20} (u,\,v)  \rightarrow  \biggl({a(C-1) \over
b_2(C-B)},\, {a(1-B) \over c_2(C-B)}\biggr), \quad  t \rightarrow
\infty. \ee  The  restriction $\beta_1=\frac{b_1-b_2}{c_2-c_1}>0$
must also be satisfied  (see (\ref{4-12})), which guarantees that
solution (\ref{4-14}) is nonnegative.
Obviously,  formula (\ref{4-20}) implies either the relation
\be\label{4-18*} B> A=1 >C \ee or the relation \be\label{4-19*}
 C > A=1 >B.\ee
 The exact solution    (\ref{4-14}) possessing  property  (\ref{4-20})
 describes the
case of a `soft' competition between two
 populations  that predicts  an arbitrarily long (in
time) coexistence of  the species $u$  and $v$.

We emphasize that all the exact solutions derived in
\cite{rod-mimura-2000,rod-mimura-2001,ch-du-04,hung12,kudrya-15,ch-dav-2011,ch-dav-2021}
 are not applicable for the description of the prey-predator interaction.
 It turns out  that the sign restrictions for the parameters
 $a_1, a_2, c_1$  and $b_2$ in  (\ref{4-9}) (see the corresponding signs in the classical system
 (\ref{0-1})) do not  guarantee positivity of the traveling fronts
 derived in the papers cited above. Motivated by this fact, we were
 able to construct an absolutely {\it new example} of a traveling front
 for the DLV system  describing  the prey-predator
 interaction of two populations.
In fact,
 using the  tanh-method \cite{malfliet-hereman-96,malfliet-04}, the
 exact solution
 \be\label{4-22}\ba u(t,x)=\frac{3a_1+a_2}{2(3b_1+b_2)}\Big[1+\tanh\Big(\sqrt{\frac{a_1b_2-a_2b_1}{8(3b_1+b_2)}}
 (x-\alpha t)\Big)\Big],\medskip\\
 v(t,x)=\frac{a_1b_2-a_2b_1}{4c(3b_1+b_2)}\Big[1+\tanh\Big(\sqrt{\frac{a_1b_2-a_2b_1}{8(3b_1+b_2)}}
 (x-\alpha t)\Big)\Big]^2,\ea\ee
 of the DLV system
 \be\label{4-23}\ba u_t=u_{xx}+u(a_1-b_1u-cv),\medskip\\
 \lambda v_t=v_{xx}+v(-a_2+b_2u-3cv)
 \ea\ee
 was discovered.
Here the restrictions
\[a_i>0, \ b_i>0, \ c>0, \ \lambda=\frac{a_2(5b_1+b_2)-2a_1b_2}{a_2b_1-3a_1(2b_1+b_2)}>0, \
\alpha=\frac{a_2b_1-3a_1(2b_1+b_2)}{\sqrt{2(3b_1+b_2)(a_1b_2-a_2b_1)}}\]
should hold.
 In the DLV system (\ref{4-23}), all parameters
are assumed to be positive. Thus, solution (\ref{4-22}) of the DLV
system (\ref{4-23}) can describe the prey-predator interaction.
Since $a_1b_2-a_2b_1>0$ (otherwise the component $v$ is negative),
we immediately obtain the restriction $\alpha<0$.
With $a_1b_2-a_2b_1>0$ and  $\alpha<0$, the following asymptotical
behavior of the above solution is obtained:
 \[
(u,\,v) \rightarrow \biggl(\frac{3a_1+a_2}{3b_1+b_2},\,
\frac{a_1b_2-a_2b_1}{c(3b_1+b_2)}\biggr), \quad t \rightarrow
\infty. \]  Such a behavior predicts  an arbitrarily long (in time)
coexistence of the preys  $u$ and the predators $v$.

In order to finish this part about traveling fronts of the
two-component DLV systems, we would like to point out the following.
  Theorems on the existence of
solutions of the Neumann problem for   DLV systems describing
competition of two species (cells, chemicals, etc.)
 have been known for a long time
(see, e.g., \cite{3-lou-ni} and works cited therein) and new
publications with pure mathematical results are published on regular
basis (see, e.g., \cite{alha-qu-19, lam-20}). In particular, it has
been established that the coefficient relations (\ref{4-18}),
(\ref{4-19}), (\ref{4-18*}) and (\ref{4-19*}) play a key role in
behavior of any solutions of (\ref{4-9}). However,
 those papers typically  do not present  such solutions in an
  explicit form. Theorem \ref{4-t1} and the above discussion show
  such solutions in the closed form. Moreover the traveling waves
  presented here satisfy no-flux conditions (the zero Neumann conditions)  at
  infinity.

Now we present some information about traveling fronts of the
three-component DLV systems.
 In contrast to the two-component DLV systems, there are very
few papers \cite{hung11,chen-hung12,chen-hung16}
    devoted to the search for   traveling wave solutions
     of the three-component DLV systems.

Probably    traveling waves  of the DLV system (\ref{3-3})  were for
the first time   identified in \cite{hung11}. Those solutions were
constructed under essential parameter restrictions. In particular,
assuming that $\lambda_1=\lambda_2=\lambda_3=1$ in (\ref{3-3}), i.e.
diffusivities of all populations are the same, the traveling wave
solution has the form  \cite{hung11}
\[\ba u(t,x)=\left(2+\alpha-\frac{a}{4}\rg)\Big[1-\tanh\Big(x-\alpha t\Big)\Big]^2,
\medskip\\
v(t,x)=\frac{a}{4}\Big[1+\tanh\Big(x-\alpha t\Big)\Big]^2,\medskip\\
w(t,x)=(a-2-\alpha)\Big[1-\tanh\Big(x-\alpha t\Big)\Big],\ea\]
provided
\[\ba  a_1=a_2=a_3=a, \ b_1=-1,
\ b_2=\frac{a-24}{8-a+4\alpha}, \ b_3=\frac{a-4-2\alpha}{8-a+4\alpha}, \medskip\\
c_1=\frac{4\alpha-a-16}{a}, \ c_2=-1, \ c_3=\frac{2\alpha-a-4}{a}, \
e_1=\frac{a-4-2\alpha}{2+\alpha-a}, \
e_2=\frac{a-4+2\alpha}{2+\alpha-a}, \ e_3=-1, \ea\] where $a$ and
$\alpha$ are arbitrary parameters. Obviously, the inequalities
$\alpha+2<a< 4(\alpha+2)$  should hold in order to guarantee
positivity of the components $u, \ v$ and $w$.  The above solution
can be  treated
as a generalization of the traveling waves
(\ref{4-1}) and (\ref{4-2}). Note that in  \cite{hung11} the
traveling wave solution for arbitrary diffusion coefficients was
constructed.

More interesting  traveling wave solutions  of the DLV system
(\ref{3-3})  were derived in \cite{chen-hung12}. In particular, by
setting the parameters as follows
\be\label{4-25}\ba\lambda_1=\lambda_2=\lambda_3=1, \ a_1=a_2=a_3=a,
\ b_1=-1,
 \ b_2=\frac{8+3a+e(24-3a)}{a(e-1)}, \ b_3=\frac{2(a+8e-a e)}{a(e-1)}, \medskip\\
c_1=\frac{8(1-3e)}{a(e-1)}, \ c_2=-1, \ c_3=\frac{8(1-3e)}{a(e-1)},
\ e_1=-e, \ e_2=\frac{(a-24)(1-e)}{16}, \ e_3=-1, \ea\ee the
traveling wave
\be\label{4-26}\ba u(t,x)=\frac{a}{2}\Big[1+\tanh\Big(x-\alpha t\Big)\Big],\medskip\\
v(t,x)=\frac{a}{4}\Big[1-\tanh\Big(x-\alpha t\Big)\Big]^2,\medskip\\
w(t,x)=\frac{4}{e-1}\Big[1-\tanh^2\Big(x-\alpha t\Big)\Big]\ea\ee
were obtained.  Here  $\alpha=\frac{a-4+20e-ae}{2(e-1)}.$ In order
to guarantee positivity of the components $u, \ v$ and $w$, the
inequalities $e>1$  and $a>0$  should hold.

\begin{figure}[h!]
\begin{center}
\includegraphics[width=7cm]{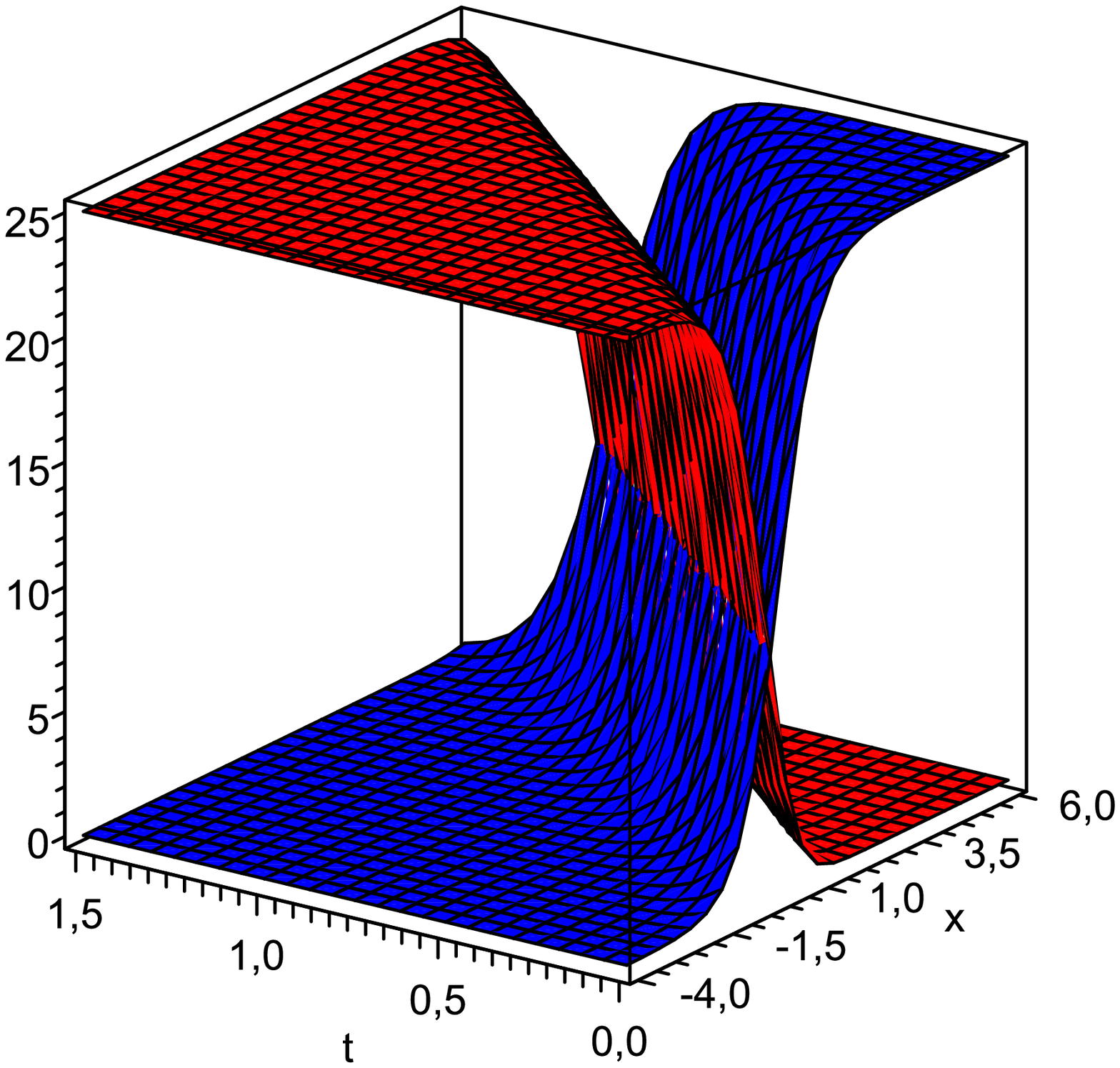}
\includegraphics[width=7cm]{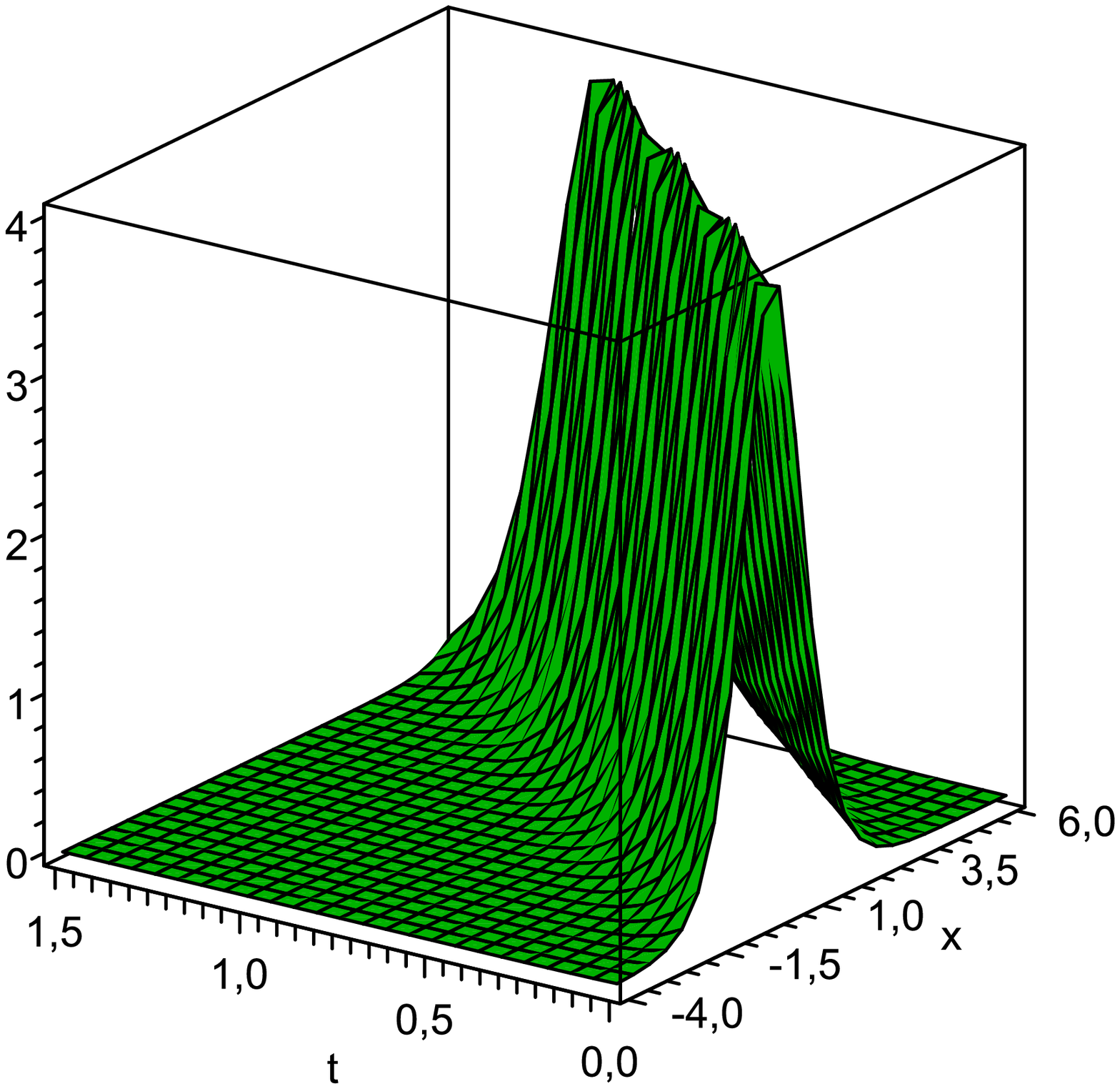}
\end{center}
\caption{Surfaces representing the components $u$ (blue), $v$ (red) and $w$ (green) of solution (\ref{4-26})
 with $a=25, \ e=2, \ \alpha=\frac{11}{2}$
 of the DLV system (\ref{3-3}) with the parameters defined by formulae (\ref{4-25}).}
\label{4-fig1}
\end{figure}

Restrictions (\ref{4-25}) define the signs of the parameters of
(\ref{3-3}). Depending on the parameter signs (\ref{3-3}) can
describe different type of interactions of species. It may also
happen that formulae  (\ref{4-25}) lead to a system, which  is not
applicable for modeling any interaction.
 Here we present an example when the exact solution can be useful.
 It can be noted that  the following additional restrictions on
parameters $a$ and $e$:
\[a>24, \quad  e>1, \quad   a(e-1)> 8e +\frac{8}{3} \]
lead to negative $b_i, \ c_i, \ e_i$ in the DLV system (\ref{3-3}).
Thus, we conclude that the system models   competition between three
populations and  the traveling fronts (\ref{4-26}) describe their
densities in time and space. Interestingly, the competition predicts
extinction of the population $w$ while other two population will
survive or die out depending on the sign of the velocity $\alpha$.
In Fig.~\ref{4-fig1}, three surfaces are presented for the exact
solution (\ref{4-26}) for the correctly-specified parameters
satisfying the above restrictions. As one concludes from
Fig.~\ref{4-fig1}, the solution describes the competition, which
leads to  the extinction of the populations $u$  and $w$, while the
population $v$ dominates as $t \to \infty$.

Finally, we present  {\it new traveling waves}  that describe
another type of  interaction between  three populations (cells,
chemicals). Assuming that  the $u$ and $v$ species compete for the
same resources and  $w$  is a predator for the above two species,
one arrives at the DLV system
\be\label{4-28}\ba \lambda_1 u_t =  u_{xx}+u(a_1-b_1u-c_1v-e_1w),\\
\lambda_2v_t = v_{xx}+ v(a_2-b_2u-c_2v-e_2w),\\
\lambda_3w_t= w_{xx}+w(-a_3+b_3u+c_3v-e_3w),\ea \ee where all
parameters are positive. The competition-prey-predator model
(\ref{4-28}) differs essentially from those studied in
\cite{hung11,chen-hung12,chen-hung16}  and can be thought as a
generalization of the two-component model (\ref{4-23}). Applying the
tanh-method, we found the traveling wave solutions of (\ref{4-28})
with correctly-specified parameters. Omitting awkward calculations,
we present only a result. Thus, the competition-prey-predator model
(\ref{4-28}) with the parameters
\[\ba  \lambda_1=\frac{2(4+a_1)}{16-a_3}, \ \lambda_2=\frac{2(4+a_2)}{16-a_3},
 \ \lambda_3=1, \medskip \\
e_1=1, \  e_2=1, \ e_3=3, \
 \ea \]
 has  traveling wave solutions of the form
\[\ba
u(t,x)=\frac{(8-a_2)c_3+(24+a_3)c_2}{2(b_3c_2-b_2c_3)}\left(1+\tanh\left(x+\frac{a_3-16}{4}\,t\right)\right),\medskip \\
v(t,x)=\frac{(a_2-8)b_3-(24+a_3)b_2}{2(b_3c_2-b_2c_3)}\left(1+\tanh\left(x+\frac{a_3-16}{4}\,t\right)\right),\medskip \\
w(t,x)=2\left(1+\tanh\left(x+\frac{a_3-16}{4}\,t\right)\right)^2,\ea
\]
provided  the restriction on  the parameters $a_k, \ b_k$ and $c_k$
($k=1,2,3$)
\[ (24+a_3)(b_1c_2-b_2c_1)= (8-a_1)(b_2c_3-b_3c_2)   +  (8-a_2)(b_3c_1-b_1c_3)\]
 holds.

Obviously, there is an infinity number of parameter sets, which
satisfy the above restriction and guarantee that the components $u,
\ v$ and $w$ are nonnegative. For instance, setting
\[a_1=11, \ a_2=9, \ a_3=4, \ b_1=\frac{1}{2}, \ b_2=\frac{1}{6}, \ b_3=5, \ c_1=6, \ c_2=2, \ c_3=7,\]
one obtains $\lambda_1=\frac{5}{2}$ and $\lambda_2=\frac{13}{6}$ and the exact
solution
\[\ba
u(t,x)=\frac{147}{53}\left(1+\tanh\left(x-3\,t\right)\right),\medskip \\
v(t,x)=\frac{1}{53}\left(1+\tanh\left(x-3\,t\right)\right),\medskip \\
w(t,x)=2\left(1+\tanh\left(x-3\,t\right)\right)^2.\ea
\]
Notably, this solution predicts that all species die out  as $t \to
\infty$.

\section{\bf Conditional symmetries of the DLV systems}\label{sec-5}

Now we turn to analysis of $Q$-conditional symmetries of the DLV
systems. It will be demonstrated that application of such symmetries
for solving  the DLV systems is more efficient
in comparison with  the Lie symmetries. First of all we note that
(\ref{1-2}) is a system of evolution equations. It is well-known
that the problem of constructing its $Q$-conditional symmetries for
systems of evolution equations  essentially depends on the function
$\xi^0$ in (\ref{2-1}) (see, e.g., \cite{ch-dav-2021}). Thus, one
should consider two different cases\,:
\begin{enumerate}
  \item $\xi^0\neq0.$
  \item $\xi^0=0, \ \xi^1\neq0.$
\end{enumerate}
In \textbf{\emph{Case\,1}},  one may set $\xi^0=1$ without loss of
 generality
applying the well-known property  of $Q$-conditional symmetry operators
(see, e.g., Section 1.2 in \cite{ch-dav-book}).
Moreover, in this case the differential consequences of equations
$Q(u^i)=0$ (that are presented in the manifold ${\cal{M}}_m$) w.r.t.
the independent  variables $t$ and  $x$ lead to  second-order PDEs
involving the  derivatives $u^i_{tt}$ and the mixed derivatives
$u^i_{tx}$. However, $u^i_{tt}$ and $u^i_{tx}$  do not occur in the
invariance conditions (\ref{2-3}). Thus, the manifold ${\cal{M}}_m$
can be rewritten as $\{S_i=0,Q(u^i)=0,i=1,\ldots,m\}$, i.e. the
first-order differential consequences can be omitted.

 It is well-known  that the task of constructing the $Q$-conditional symmetries
 in \textbf{\emph{Case\,2}} ($\xi^0=0$) for scalar evolution equations
  is equivalent to solving  the equation in question \cite{zhdanov}.
  This statement can be extended on evolution systems of PDEs. In other words,
  it means that application of the invariance criteria (\ref{2-3}) to operator
   (\ref{2-1}) with $\xi^0=0$ after cumbersome calculations leads to a system of DEs,
   which is equivalent to (\ref{1-2}). So, in the case of nonlinear and nonintegrable
   equations (systems), one can identify only some \emph{particular cases}
   of the $Q$-conditional symmetry operators of the form (\ref{2-1}) with
$\xi^0=0$.
   In  \cite{ch-dav-2021}, the two-component DLV system (\ref{0-4}) was examined
   in order to find $Q$-conditional symmetries in \textbf{\emph{Case\,2}}
   (the so-called no-go case) using Definition~\ref{d2}.

The system of DEs for finding $Q$-conditional symmetries of the DLV
system (\ref{3-1}) for the first time was derived
    in  \cite{ch-dav-2011}. An algorithm based on
    Definition~\ref{d1}  was applied for this purpose.
    In particular, it was shown that the structure of any  $Q$-conditional
    symmetry of (\ref{3-1})  can be specified as follows
    \be\label{c68}
Q=\p_t+\xi\p_x+\lf(q^1v+r^1u+p^1\rg)\p_u+\lf(q^2u+r^2v+p^2\rg)\p_v,\ee
where the functions $q^k(t,x),\ r^k(t,x)$ and $ p^k(t,x) \ (k=1,2)$
should be found from the remaining equations of the system of DEs
(see equations (30)--(45) in \cite{ch-dav-2011}). The system is very
    complicated and was not  completely integrated, however
    important  results were derived.
In particular,  the following  existence theorem  was proved.
 \begin{Theorem}
\label{5-t1*} \cite{ch-dav-2011} In the case $\lambda_1=\lambda_2$,
DLV system (\ref{3-1}) admits only such  $Q$-conditional operators
of the form (\ref{c68}),
 which are
equivalent to the Lie symmetry operators. In the case
$\lambda_1\neq\lambda_2$,  DLV system (\ref{3-1})  is
$Q$-conditionally invariant under operator (\ref{c68}) if and only
if $b_1=b_2=b$  and $ c_1=c_2=c.$ \end{Theorem}

In order to find symmetries in explicit forms, the case
$\lambda_1\neq\lambda_2$ was examined. As a result, the following
theorem was proved.

\begin{Theorem} \label{5-t1}\cite{ch-dav-2011} If  $bc=0,\ b^2+c^2\neq0$ then
system  (\ref{3-1}) and the $Q$-conditional symmetries (up to the
transformations $u\rightarrow bu, \ v\rightarrow
\exp\lf(\frac{a_2}{\lambda_2}t\rg)v, \ b\neq0$  and $u\rightarrow
\exp\lf(\frac{a_1}{\lambda_1}t\rg)v,\ cv\rightarrow u,$ $c\neq0$)
have the forms
\[\begin{array}{l} \lbd_1 u_t=
u_{xx}+u(a_1+u),\\ \lbd_2 v_t= v_{xx}+ vu,\medskip\\
Q=\p_t+\frac{2\al_1}{\lbd_1-\lbd_2} \,
\p_x+\lf(\varphi(t)\exp(\alpha_1x)u+
\exp(\alpha_1x)\lf(\lambda_2\varphi'(t) +a_1\varphi(t)-
\alpha^2_1\varphi(t)\rg)+\al_2v\rg)\p_v,
\end{array}\] where
 the function $\varphi(t)\neq0$ is  the general
solution of the linear ODE
\[\lambda^2_2\varphi''+
\lambda_2(a_1-2\alpha^2_1)\varphi'+\alpha^2_1(\alpha^2_1-a_1)\varphi=0.\]

If  $bc\neq0$ and the additional restrictions
\be\label{5-2}q^1_x=q^2_x=0\ee hold then exactly three cases (up to
the transformations $u\rightarrow bu, \ v\rightarrow cv $ and
$u\rightarrow v, \ v\rightarrow u $) exist when   system (\ref{3-1})
is invariant w.r.t. $Q$-conditional symmetry operators. These cases
are listed in Table~\ref{5-tab3} \end{Theorem}

\begin{table}[h!]
\caption{$Q$-conditional symmetries of the DLV system (\ref{3-1})
with $\lambda_1\neq\lambda_2$ and $b_1=b_2=b, \ c_1=c_2=c, \
bc\neq0$.} \label{5-tab3}
\begin{small}
\begin{center}
\begin{tabular}{|c|c|c|c|}
\hline  & DLV systems &  Restrictions & Operators  \\  \hline &&&\\
 1 &
$\lambda_1u_t = u_{xx}+u(a_1+u+v)$ & $a_1\neq a_2$& $(\lbd_1-\lbd_2)\p_t-(a_1v+
a_2u+a_1a_2)(\p_u-\p_v), \ a_1a_2\neq0;$  \\
& $\lambda_2v_t = v_{xx}+ v(a_2+u+v)$& &$(\lbd_1-\lbd_2)\p_t+(a_1-a_2)u(\p_u-\p_v);$
\\ &&&$(\lbd_1-\lbd_2)\p_t-(a_1-a_2)v(\p_u - \p_v)$
  \\  \hline &&&\\

 2 &
$\lambda_1u_t = u_{xx}+u(a+u+v)$ &
 & $(\lbd_1-\lbd_2)\p_t-a(v+
u+a)(\p_u-\p_v), \ a\neq0;$
   \\  & $\lambda_2v_t = v_{xx}+ v(a+u+v)$&&$(\lbd_1-\lbd_2)t\p_t-(\lambda_1v+\lambda_2u)(\p_u-\p_v)$
 \\ \hline &&&\\

 3 &
$\lbd_1 u_t = u_{xx}+u(a\lambda_1+u+v)$
  & $a\neq0$&$(\lbd_1-\lbd_2)\p_t-a(\lbd_1v+
 \lbd_2u+a\lbd_1\lbd_2)(\p_u-\p_v);$
  \\ &$\lbd_2 v_t = v_{xx}+ v(a\lambda_2+u+v)$&&$\p_t+au(\p_u-\p_v);\
\p_t-a v(\p_u-\p_v);$ \\ &&&$\p_t+\frac{a\alpha(\lbd_1v+
 \lbd_2u+a\lbd_1\lbd_2)}{(e^{-at}-\alpha(\lbd_1-\lbd_2))}\,(\p_u-\p_v), \alpha\neq0$
  \\ \hline
\end{tabular}
\end{center}
\end{small}
\end{table}

\begin{Remark} In contrast to Definition~\ref{d1}, applying Definition~\ref{d2}
 leads to a complete description (i.e. without  additional restrictions)
of $Q$-conditional symmetries of the first type (with $\xi^0\neq0$)
of the DLV system (\ref{3-1})   (see Theorem~2~\cite{ch-dav-2011}).
Unfortunately  all the $Q$-conditional symmetries of the first type,
which have been derived, coincide with those listed in
Table~\ref{5-tab3}.
\end{Remark}

Now we turn to \textbf{\emph{Case\,2}}, i.e. the no-go case
($\xi^0=0$, see operator  (\ref{2-1})), which was   was investigated
    in \cite{ch-dav-2021}. As  mentioned above, the
    algorithm based on Definition~\ref{d1} leads to an unsolvable
    system of DEs in this case. So, we used Definition~\ref{d2}
     in order to find all possible  $Q$-conditional symmetries of the first type.
      The  main result  can be formulated as follows.

\begin{Theorem}\label{5-t2}\cite{ch-dav-2021} The DLV system  (\ref{3-1})  with restrictions (\ref{3-6}) is invariant under $Q$-conditional symmetry operator(s)  of the first
type \be\label{5-3} Q = \
\xi(t,x,u,v)\partial_x+\eta^1(t,x,u,v)\partial_u+\eta^2(t,x,u,v)\partial_v,
\ \xi\neq0, \ee if and only if the system  and the relevant
operator(s)
are as specified  in Table~\ref{5-tab1}. Any other DLV system
(\ref{3-1}) admitting a  $Q$-conditional symmetry of the first type
and the corresponding operator(s) are reducible to those listed in
Table~\ref{5-tab1} by an appropriate  transformation from the set
 \[t^*=t+t_0, \ x^*=e^{\gamma_0}(x+ x_0),  \ u^*=\beta_{11}\,e^{\gamma_1t}u+\beta_{12}\,v, \
  v^*=\beta_{22}\,e^{\gamma_2t}v+\beta_{21}\,u,
  \] where $t_0, \  x_0, \ \beta_{ij}$ and $\gamma_j$ are  correctly-specified constants.
\end{Theorem}

\begin{table}[h!]
\caption{$Q$-conditional symmetries of the first type of the DLV system
(\ref{3-1})}
\label{5-tab1}
\begin{small}
\begin{center}
\begin{tabular}{|c|c|c|c|}
\hline  & DLV systems &  Restrictions & Operators  \\  \hline &&&\\
 1 &
$\lambda_1u_t = u_{xx}+u(a_1+u+v)$ & $\lambda_1\neq \lambda_2$& $Q^u_1=\partial_x+\frac{g^1_x}{g^1}\,u\,(\partial_u-\partial_v),$  \\
& $\lambda_2v_t = v_{xx}+ v(a_2+\frac{\lambda_2}{\lambda_1}\,u+\frac{\lambda_2}{\lambda_1}\,v)$& &$Q^v_1=\partial_x+\frac{g^2_x}{g^2}\,v\,(\partial_v-\partial_u)$  \\  \hline &&&\\

 2 &
$u_t = u_{xx}+ u(a+u+2v)$ &
 & $Q^v_2=G(x,v)\left(\partial_x+F(x,v)(\partial_u-\partial_v)\right)$
   \\  & $\lambda v_t =  v_{xx}+v(a+v)$&&
 \\ \hline &&&\\

 3 &
$u_t = u_{xx}+uv$
  & $a_2c_2\neq0$&$Q^{u}_3=\partial_x+r(t,x)\,u\,\partial_u,$
  \\ &$\lambda v_t = v_{xx}+ v(a_2+c_2v)$&&$Q^{v}_3=\left(h^1(\omega)-2th^2(\omega)\right)\partial_x$ \\ &&&$+\left((h^2(\omega)x+h^3(\omega))u+p(t,x,v)\right)\partial_u$
  \\ \hline &&&\\

4 & $u_t =  u_{xx}+uv$
 & $c_2\neq0$&$Q^{u}_3 , \  Q^v_4=\left(h^1(\theta)-2th^2(\theta)\right)\partial_x$
 \\ &$\lambda v_t = v_{xx}+ c_2v^2$&&$+\left((h^2(\theta)x+h^3(\theta))u+p(t,x,v)\right)\partial_u$
\\ \hline &&&\\

5 & $u_t = u_{xx}+uv$
 &  $ a_2\neq0$&
$Q^{u}_3 , \quad Q^{v}_3, \quad  Q^v_5=\partial_x+e^{a_2t}u\partial_v$
 \\ &$v_t = v_{xx}+ v\left(a_2+\frac{v}{2}\right)$&&$+\left(\alpha\, u-\frac{e^{-a_2t}}{2}\,v^2-a_2e^{-a_2t}v\right)\partial_u$
\\ \hline &&&\\

6 & $u_t =  u_{xx}+uv$
 & &  $Q^{u}_3 , \ Q^v_4,$
 \\ &$v_t = v_{xx}+ \frac{1}{2}\,v^2$&&$
Q^v_6=(\alpha_1t+\alpha_0)\partial_x+
(\alpha_1t+\alpha_0)u\partial_v$ \\ &&&$+\left(\left(\alpha_2-\frac{\alpha_1}{2}\,x\right)u-\frac{\alpha_1t+\alpha_0}{2}\,v^2-\alpha_1v\right)\partial_u$
\\  \hline &&&\\

7 &
$u_t = u_{xx}+uv$
 &$a_2\neq0$&$ \alpha_1^2+\alpha_2^2\neq0$, \  $Q^{u}_3, \  Q^{v}_3, $
  \\ &  $v_t = v_{xx}+ v(a_2+v)$&&$ Q^u_7=\partial_x+\left(-\frac{x}{2t}\,u+\frac{\alpha_1}{t}\right.$ \\ &&&$+\left.\left(\frac{\alpha_2e^{-a_2t}}{t}+\frac{\alpha_1}{a_2t}\right)v\right)\partial_u$
   \\ \hline &&&\\

8 & $u_t =  u_{xx}+uv$
&$ \alpha_1^2+\alpha_2^2\neq0$& $Q^{u}_3, \  Q^v_4, $
 \\ &$v_t = v_{xx}+ v^2$&& $ Q^u_8=\partial_x+\left(-\frac{x}{2t}\,u+\frac{\alpha_1}{t}+\left(\frac{\alpha_2}{t}+\alpha_1\right)v\right)\partial_u$
 \\ \hline
\end{tabular}
\end{center}
\end{small}
\end{table}

\begin{Remark}In Table~\ref{5-tab1}, the upper indices $u$ and $v$ mean that the relevant $Q$-conditional symmetry operators satisfy Definition~\ref{d2} in the case of the manifold
${\cal{M}}^1_1$ ($u^{1}=u$) and ${\cal{M}}^2_1$  ($u^{2}=v$), respectively.
\end{Remark}
\begin{Remark}
In Table~\ref{5-tab1}, $\omega=\frac{a_2+c_2v}{\lambda
v}\,e^{\frac{a_2}{\lambda}\,t}, \ \theta=t+\frac{\lambda}{c_2v};$
$h^1, \ h^2$ and $h^3$ are arbitrary smooth functions of the
relevant variables, while the function $p(t,x,v)$ is the general
solution of the linear ODE
 \[p_t=p_{xx}-\frac{v(a_2+c_2v)}{\lambda}p_v+vp,\]
 the functions $F$ and $G$ form the general solution of the system
\[ FF_v-F_x+av+v^2=0, \ G_x=FG_v,\]
 the function $r(t,x)$ is the general solution
of the Burgers equation \[r_t=r_{xx}+2rr_x,\] while
\begin{equation}\label{5-6} g^i(t,x)=\left\{
\begin{array}{l} \alpha_0\exp\left(\frac{\kappa^2t}{\lambda_i}\right)+\alpha_1\sin(\kappa\, x)+\alpha_2\cos(\kappa\, x), \quad
   \mbox{if} \quad \frac{\lambda_1a_2-\lambda_2a_1}{\lambda_1-\lambda_2}>0,
 \medskip \\ \alpha_0\exp\left(-\frac{\kappa^2t}{\lambda_i}\right)+\alpha_1e^{\kappa x}+\alpha_2e^{-\kappa x}, \quad \mbox{if} \quad \frac{\lambda_1a_2-\lambda_2a_1}{\lambda_1-\lambda_2}<0,
 \medskip  \\
 \alpha_0+\alpha_1x+\alpha_2\lambda_ix^2+2\alpha_2t, \quad \mbox{if} \quad \lambda_1a_2=\lambda_2a_1,
\end{array} \right.
\end{equation} where $i=1,2,$ $\kappa=\sqrt{\left|\frac{\lambda_1a_2-\lambda_2a_1}{\lambda_1-\lambda_2}\right|},$ $\alpha_0, \ \alpha_1$ and $\alpha_2$ are arbitrary constants.
\end{Remark}

It should  be noted that all the systems arising in
Table~\ref{5-tab1}, except that in Case~1, are semi-coupled (see the
second equation is each system).
 We point out  that the second equation in Cases~2, 3, 5 and 7
  is nothing else but the famous Fisher equation. Obviously,
  Case~1 from Table~\ref{5-tab1} is the most interesting from
  applicability point of view. In fact,   it will be demonstrated in
  Section~\ref{sec-6} that the  system from Case~1 models  competition  of two
  populations of species (cells, chemicals) and the relevant exact
  solutions will be constructed.

Now we turn to the three-component DLV system.
A complete description of the $Q$-conditional (nonclassical)
symmetry for the three-component DLV system is still unknown from
the same reason as for the two-component  system.
 To the best of our knowledge, there is only a single  study
\cite{ch-dav2013}  devoted only to the search for conditional
symmetries of the three-component DLV system. In that paper, all
possible  $Q$-conditional symmetries of the first type were derived
in \textbf{\emph{Case\,1}}.

\begin{Theorem}\label{5-t3} \cite{ch-dav2013} The DLV system  (\ref{3-3})
    is invariant under $Q$-conditional symmetry
of the first type
\[\begin{array}{l}
Q = \xi^0 (t, x, u, v, w)\p_{t} + \xi^1 (t, x, u, v, w)\p_{x} +  
 \eta^1(t, x, u, v, w)\p_{u}+\\ \hskip1.8cm \eta^2(t, x, u, v, w)\p_{v}+\eta^3(t, x, u, v, w)\p_{w}, \ \xi^0\neq0,
  \end{array}\]
if and only if  it  and the relevant operators are as specified  in
Table~\ref{5-tab2}. Any other DLV system admitting a $Q$-conditional
symmetry operator of the first type is reduced to one of those from
Table~\ref{5-tab2} by a transformation from the set  (\ref{3-5}).
\end{Theorem}

\begin{table}
\caption{$Q$-conditional symmetries  of the first type of the  DLV system  (\ref{3-3})}
\label{5-tab2}
\begin{small}
\begin{center}
\begin{tabular}{|c|c|c|c|} \hline
& Reaction terms &Restrictions & $Q$-conditional symmetry operators \\  \hline &&& \\
1 &
$u(a_1+bu+bv+ew)$   &$(b-1)^2+(e-e_3)^2\neq0,$ & $\p_t+\frac{a_1-a_2}{\lambda_1-\lambda_2}u(\p_u-\p_v),$
 \\ &$v(a_2+bu+bv+ew)$&$a_1\neq a_2$&$\p_t+\frac{a_1-a_2}{\lambda_1-\lambda_2}v(\p_v-\p_u)$ \\ &$w(a_3+u+v+e_3w)$&&
  \\  \hline &&&\\
 2 & $u(a_1+u+v+w)$ &$(a_1-a_2)^2+(a_1-a_3)^2\neq0$  &
$Q^2_i, \ i=1,\dots,6$
 \\ &$v(a_2+u+v+w)$&& \\ &$w(a_3+u+v+w)$&&
  \\  \hline &&&\\
3 &  $u(a_1+u+v+w)$
&$(\lambda_2-\lambda_3)a_1-\lambda_2a_3$  & $Q^2_i, \ i=1, \dots , 6,$
 \\ &$v(a_2+u+v+w)$&$+\lambda_3a_2=0,$&
 $\p_t+\beta\exp\lf(\frac{a_2-a_3}{\lambda_2-\lambda_3}t\rg)u(\p_v-\p_w)$ \\ &$w(a_3+u+v+w)$&$a_2\neq a_3, \ \beta\neq0$&
  \\  \hline &&& \\
 4 & $u(a_1+u+v+w)$ &
$(\lambda_2-\lambda_3)a_1-(\lambda_1-\lambda_3)a_2$&
$Q^4_i, \ i=1,\dots,6$
 \\ &$v(a_2+u+v+w)$&$+(\lambda_1-\lambda_2)a_3=0,$& \\ &$w(a_3+u+v+w)$&$(a_1-a_2)^2+\alpha^2\neq0 $&
  \\  \hline &&&\\
5 & $u(a_1+bu+v)$ & $(b-1)^2+(c-1)^2\neq0$ & $Q^5_1$
 \\ &$v(a_2+u+cv)$&& \\ &$w(bu+v)$&&
  \\  \hline &&&\\
6 & $u(a_1+u+v)$   & &
  $Q^5_1, \ Q^6_i, \ i=1,\dots,4$
 \\ &$v(a_2+u+v)$&& \\ &$w(u+v)$&&
  \\  \hline &&&\\
 7 &$u(a_1+bu+cv)$
&
$\lambda_2=\lambda_3=1, \ b\neq1, \ c\neq1,$& $\p_t+\lf((1-b)u+(1-c)v+a_2(1-c)\rg)\p_w$
 \\ &$v(a_2+u+v)$&$
a_1(1-b)=a_2b(1-c)$& \\ &$w(bu+v)$&&
  \\  \hline &&&\\
8 &   $u(a+bu+cv)$
  & $\lambda_2=\lambda_3=1,
\ b\neq1, \ c\neq1,$&
$\p_t+(1-c)\p_w$
 \\ &$v(a+u+v)$&$b(2-c)=1$&$+\lf((1-b)u+(1-c)v\rg)\varphi_4(t)\p_w$ \\ &$w(bu+v)$&&
  \\  \hline &&&\\
9 &   $u(a_1+u+v)$  &
$\lambda_2=\lambda_3=1$&
$Q^9_i, \ i=1, \dots , 5$
 \\ &$v(a_2+u+v)$&& \\ &$w(u+v)$&&
  \\  \hline
\end{tabular}
\end{center}
\end{small}
\emph{The coefficients $\lambda_k>0 \ (k=1,2,3)$ are assumed to be
different in cases 1--6.}
\end{table}

In Table~\ref{5-tab2}, the following designations are
introduced:
\[Q^2_i=Q^4_i \ \mbox{with} \  \alpha=0, \ i=1, \dots, 6;\]
\[ Q^4_1=\p_t+\frac{a_1-a_2}{\lambda_1-\lambda_2}u(\p_u-\p_v)+\alpha u(\p_v-\p_w), \
Q^4_2=\p_t+\frac{a_1-a_2}{\lambda_1-\lambda_2}v(\p_v-\p_u)+\alpha
v(\p_u-\p_w),\]
\[Q^4_3=\p_t+\frac{a_1-a_3}{\lambda_1-\lambda_3}u(\p_u-\p_w)+\alpha
u(\p_v-\p_w), \  Q^4_4=\p_t+\frac{a_1-a_3}{\lambda_1-\lambda_3}w(\p_w-\p_u)+\alpha
w(\p_u-\p_v)
 , \]
\[Q^4_5=\p_t+\frac{a_2-a_3}{\lambda_2-\lambda_3}v(\p_v-\p_w)+\alpha
v(\p_u-\p_w), \ Q^4_6=\p_t+\frac{a_2-a_3}{\lambda_2-\lambda_3}w(\p_w-\p_v)+\alpha
w(\p_u-\p_v) ;\]
\[Q^5_1=\p_t+\alpha_1\p_x+ \exp\lf(\lf(\frac{(\lambda_1-\lambda_3)^2}{4}\alpha^2_1-
a_1\rg)\frac{t}{\lambda_3}+\frac{\lambda_1-\lambda_3}{2}\alpha_1x\rg)u\p_w;
 \]
 \[Q^6_1=\p_t+\alpha_1\p_x+\exp\lf(\lf(\frac{(\lambda_2-\lambda_3)^2}{4}\alpha^2_1-a_2\rg)\frac{t}{\lambda_3}+
\frac{\lambda_2-\lambda_3}{2}\alpha_1x\rg)v\p_w,\]
 \[Q^6_2=\p_t+\frac{a_1-a_2}{\lambda_1-\lambda_2}u(\p_u-\p_v)+
 \beta\exp\lf(\frac{(\lambda_1-\lambda_3)a_2-(\lambda_2-\lambda_3)a_1}{\lambda_3(\lambda_2-\lambda_1)}t\rg)u\p_w,\]
\[Q^6_3=\p_t+\frac{a_1-a_2}{\lambda_1-\lambda_2}v(\p_v-\p_u)+
\beta\exp\lf(\frac{(\lambda_2-\lambda_3)a_1-(\lambda_1-\lambda_3)a_2}{\lambda_3(\lambda_1-\lambda_2)}t\rg)v\p_w,\]
\[Q^6_4=\p_t+\frac{a_2\lambda_1-a_1\lambda_2}{\lambda_3(\lambda_2-\lambda_1)}w\p_w+
\exp\lf(\frac{(\lambda_3-\lambda_2)a_1-(\lambda_3-\lambda_1)a_2}{\lambda_3(\lambda_1-\lambda_2)}t\rg)w(\p_u-\p_v);\]
\[ Q^9_1=Q^5_1 \ \mbox{with} \ \lambda_3=1,
\quad Q^9_2=Q^6_4 \ \mbox{with} \ \lambda_2=\lambda_3=1,\]
\[Q^9_{3}=\p_t+\frac{a_1-a_2}{\lambda_1-1}u(\p_u-\p_v)+\lf(\varphi_1(t)u+\varphi_2(t)v+\beta_1\rg)\p_w, \]
\[Q^9_{4}=\p_t+\lf(\varphi_3(t)u+\varphi_2(t)v+\beta_1\rg)\p_w,
\quad Q^9_5=\p_t+\frac{a_1-a_2}{\lambda_1-1}v(\p_v-\p_u);\]
 where the functions $\varphi_i(t) \
(i=1,\dots,4)$  are as follows: \[ \varphi_1(t)=
\left\{
\begin{array}{l}
\beta_1t + \beta_2, \ \mbox{if} \ a_2=0,\\
\beta_2\exp(-a_2t)+\frac{\beta_1}{a_2}, \
   \mbox{if} \  a_2\neq0,
\end{array} \right.   \quad \varphi_2(t)= \left\{ \begin{array}{l}
\beta_1t, \  \mbox{if} \ a_2=0,\\
\frac{\beta_1}{a_2}, \
   \mbox{if} \  a_2\neq0,
\end{array} \right.
\]
\[\varphi_3(t)= \left\{ \begin{array}{l} \beta_1t + \beta_2, \  \mbox{if} \ a_1=0,\\
\beta_2\exp(-a_1t)+\frac{\beta_1}{a_1}, \
   \mbox{if} \  a_1\neq0,
\end{array} \right.  \quad   \varphi_4(t)= \left\{ \begin{array}{l} t + \beta, \ \mbox{if} \ a=0, \\
\beta\exp(-at)+\frac{1}{a}, \
   \mbox{if} \  a\neq0,\end{array} \right.
\]
  while  $ \alpha$ and $\beta$ (with and without subscripts 1 and 2)  are arbitrary constants.

\begin{Remark}
  The inequalities listed  in the third
   column of Table~\ref{5-tab2} guarantee  that the $Q$-conditional
    symmetries  from the fourth column are not equivalent to any
     Lie  symmetry  presented in Table~\ref{3-tab2}.
\end{Remark}

We conclude that the three-component DLV system, depending on the
coefficient restrictions,  admits a wider range of $Q$-conditional
symmetries of the first type compared to those  for the
two-component DLV system. In particular, there are cases when DLV
system (\ref{3-3}) admits  sets consisting of 5, 6 and even 7
different symmetries. All  these symmetries can be successfully used
for finding exact solutions.

From the applicability point of view, the systems arising in cases
1--4
 of Table~\ref{5-tab2} are  most promising   because their nonzero
  coefficients do not affect the biological sense of these systems.
  In the Section~\ref{sec-7}, we examine    these  systems.

Concluding this section, we present a short statement about the
$m$-component DLV system (\ref{2-1}).
  In the case of the   DLV system (\ref{2-1}) with $m>3$, the
        problem of constructing conditional symmetries  is still open.
      {\it Some particular results} can
          be obtained     by a simple
         generalization of the results obtained for three-component
          system.  In particular, we  proved that the
          $m$-component  system \cite{ch-dav2013}
            \[  \lbd_i u^i_t =  u^i_{xx}+u^i(a_i+u^1+\dots+u^m), \  i=1,2,\dots,m  \]
            admits $m(m-1)$ operators of the form
             \[  Q_{ij}=\p_t+\frac{a_i-a_j}{\lambda_i-\lambda_j}u^i\lf(\p_{u^i}-\p_{u^j}\rg),
              \ \ i \not=j =1,2,\dots,m, \]
              provided $(a_i-a_j)(\lambda_i-\lambda_j)\not=0$.
One may consider the above system and
          the operators   as a generalization of those presented
           in  Case 2 of Table~\ref{5-tab2} on the case of
         the $m$-component  DLV systems.

\section{\bf Exact solutions of the two-component DLV system}\label{sec-6}

This section is devoted to the construction of exact
 solutions with more complicated structures than the traveling wave solutions
  presented in Section~\ref{sec-4}.
It should be pointed out that the traveling wave solutions cannot be
applied for solving practical  models, in particular based on the
DLV systems, describing processes  in  bounded domains. In fact, any
traveling front does not satisfy  typical boundary conditions like
no-flux conditions or/and constant densities at a bounded interval.
It means that an ansatz of the form (\ref{1-2**}) should be used for
search for exact  solutions.

It is well-known that using $Q$-conditional symmetries a  given
two-dimensional PDE (system of PDEs) can be  reduced to an ODE
(system of ODEs) via the same algorithm as for classical Lie
symmetries. It means that the  ansatz corresponding to the given
operator $Q$ can be constructed  provided   the linear
(quasi-linear) first-order PDEs
 \be\label{6-1} {Q}(u)=0, \ {Q}(v)=0 \ee  are  solved.

Theorems~\ref{5-t1} and~\ref{5-t2} give several possibilities for
finding exact solutions of the
  DLV system with  correctly-specified coefficients.

Let us consider the DLV system from Case~1 of Table~\ref{5-tab3}, namely\,:
\be\label{5-4}\begin{array}{l}
 \lbd_1 u_t = u_{xx}+u(a_1+u+v),\\
 \lbd_2 v_t = v_{xx}+ v(a_2+u+v),  \ a_1\neq a_2,\ea\ee
and its $Q$-conditional symmetry operator
\be\label{5-5} Q=(\lbd_1-\lbd_2)\p_t-(a_1v+
a_2u+a_1a_2)(\p_u-\p_v).\ee

In the  case of operator (\ref{5-5}), system (\ref{6-1}) takes the form
\begin{equation}\label{6-2}\begin{array}{l}
(\lambda_1-\lambda_2)u_t=-(a_1v+a_2u+a_1a_2),\\
(\lambda_1-\lambda_2)v_t=a_1v+a_2u+a_1a_2. \end{array}\end{equation}
It follows  immediately from (\ref{6-2}) that $u_t=-v_t$, hence
\be\label{6-3}u(t,x)=-v(t,x)+\varphi_1(x).\ee Substituting
(\ref{6-3}) into the second equation of  (\ref{6-2}), one  obtains
the linear equation
\[(\lambda_1-\lambda_2)v_t=(a_1-a_2)v+a_2\varphi_1(x)+a_1a_2.\]
Since  $a_1\neq a_2$  this equation has the general solution
\[v=\frac{1}{a_1-a_2}\lf(\exp\lf(\frac{a_1-a_2}
{\lambda_1-\lambda_2}t\rg)\varphi_2(x)-a_2\varphi_1(x)-a_1a_2\rg),\]
therefore the ansatz
\begin{equation}\label{6-4}\begin{array}{l}
u=\frac{1}{a_1-a_2}\lf(-\exp\lf(\frac{a_1-a_2}
{\lambda_1-\lambda_2}t\rg)\varphi_2(x)+a_1\varphi_1(x)+a_1a_2\rg),\\
v=\frac{1}{a_1-a_2}\lf(\exp\lf(\frac{a_1-a_2}
{\lambda_1-\lambda_2}t\rg)\varphi_2(x)-a_2\varphi_1(x)-a_1a_2\rg)
\end{array}\end{equation} is obtained. Here  $\varphi_1$ and $\varphi_2$ are functions  to be found.

To obtain the reduced system, we substitute ansatz (\ref{6-4}) into
(\ref{5-4}). This means that we simply calculate the derivatives
$u_t, \ v_t, \ u_{xx}, \ v_{xx},$  and insert them into (\ref{5-4}).
Making   relevant calculations,  one arrives at the ODE system
\begin{equation}\label{6-5}\begin{array}{l}
\varphi''_1+\varphi^2_1+(a_1+a_2)\varphi_1+a_1a_2=0,\\
\varphi''_2+\frac{a_2\lambda_1-a_1\lambda_2}{\lambda_1-\lambda_2}\,\varphi_2+
\varphi_1\varphi_2=0 \end{array}\end{equation} to find the functions
$\varphi_1$ and $\varphi_2$.

\begin{Remark} Using the second and  third   operators  listed  in Case~1 of Table~\ref{5-tab3},
 one can obtain   reduced systems in a similar way  and look for exact solutions.
 However, we have checked that
 the ODE systems  obtained  simply follow from  (\ref{6-5}) by
 removing the terms $a_1\varphi_1+a_1a_2$.
\end{Remark}

In order to construct exact solutions, now we examine the ODE
systems obtained above.
To the best of our knowledge,  the general solution of the nonlinear
ODE system (\ref{6-5}) is unknown, therefore  we look for its
particular solutions. Setting $\varphi_1=\alpha=const,$ we conclude
\[ \alpha^2+(a_1+a_2)\alpha+a_1a_2=0 \ \Rightarrow  \ \alpha_1=-a_1, \
\alpha_2=-a_2 \] from the first equation of  system (\ref{6-5}).
 So, setting   $\varphi_1=-a_1$ (the case $\varphi_1=-a_2$ leads
 to the solution with the same structure) and substituting into
 the second
 equation of system (\ref{6-5}), we obtain the linear ODE:
 \be\label{6-6}
 \varphi''_2-\beta \lambda_1\varphi_2=0,
 \ee
where $\beta=\frac{a_1-a_2}{\lambda_1-\lambda_2}\not=0.$ Depending
on the sign of the parameter  $\beta$ the linear ODE (\ref{6-6})
possesses two families of   general solutions. These solutions and
ansatz (\ref{6-4}) lead to   the  following
 exact solutions of the DLV system (\ref{5-4}):
  \[\begin{array}{l}
u=-a_1+\frac{1}{a_2-a_1}\lf(C_1\exp\lf(\sqrt{\beta\lambda_1}x\rg)+C_2\exp\lf(-\sqrt{\beta\lambda_1}x\rg)\rg)e^{\beta
t},\\
v=\frac{1}{a_1-a_2}\lf(C_1\exp\lf(\sqrt{\beta\lambda_1}x\rg)+C_2\exp\lf(-\sqrt{\beta\lambda_1}x\rg)\rg)e^{\beta
t},\end{array} \] if $\beta>0,$  and
\begin{equation}\label{6-7}\begin{array}{l}
u=-a_1+\frac{1}{a_2-a_1}\lf(C_1\cos\lf(\sqrt{-\beta\lambda_1}x\rg)+C_2\sin\lf(\sqrt{-\beta\lambda_1}x\rg)\rg)e^{\beta
t},\\
v=\frac{1}{a_1-a_2}\lf(C_1\cos\lf(\sqrt{-\beta\lambda_1}x\rg)+C_2\sin\lf(\sqrt{-\beta\lambda_1}x\rg)\rg)e^{\beta
t}, \end{array}\end{equation} if  $\beta<0$ (hereinafter  $C_1$ and
$C_2$ are arbitrary constants).

Now we demonstrate that extra exact  solutions of  (\ref{6-5}) can
be derived  provided  some restrictions on $\lambda_1$ and
$\lambda_2$
 take place. Indeed, we note that the substitution \be\label{6-8}
\varphi_1=\varphi-a_1 \ee simplifies the first equation of
(\ref{6-5}) to the form
 \be\label{6-9}
\varphi''+\varphi^2+(a_2-a_1)\varphi=0. \ee Of course, (\ref{6-9})
can be reduced to the first-order ODE
\[\lf(\frac{d\varphi}{dx}\rg)^2=-\frac{2}{3} \
\varphi^3+(a_1-a_2)\varphi^2+C \ \] with the general solution
involving the  Weierstrass function \cite{b-e}. Now we set $C=0$ in
order to avoid cumbersome formulae,  therefore  the general solution
is \be\label{6-10}
\varphi=\frac{3(a_1-a_2)}{2}\lf(1-\tanh^2\lf(\frac{\sqrt{a_1-a_2}}{2}
\,x\rg)\rg), \ee  if  $a_1>a_2$,  and \be\label{6-11}
\varphi=\frac{3(a_1-a_2)}{2}\lf(1+\tan^2\lf(\frac{\sqrt{a_2-a_1}}{2}\,x\rg)\rg),
\ee   if $a_1<a_2.$

Thus, we can apply formulae (\ref{6-10}) and (\ref{6-11}) to solve
the second ODE of (\ref{6-5}). In the case of solution (\ref{6-10}),
this ODE takes the form  \be\label{6-12}
\varphi_2''+\varphi_2(a_1-a_2)\lf(\frac{\lambda_1-3\lambda_2}{2(\lambda_1-\lambda_2)}-
\frac{3}{2}\tanh^2\lf(\frac{\sqrt{a_1-a_2}}{2}\,x\rg)\rg)=0 .\ee The
general solution of (\ref{6-12}) with some restrictions on
$\lambda_1$ and $\lambda_2$ can be found \cite{pol-za}:
 \be\label{6-13} \varphi_2=f_1(x)\lf(C_1+C_2\int\frac{1}{f^2_1(x)}\,dx\rg),\ee if
$\lambda_1=\frac{9}{5}\lambda_2$, and
 \be\label{6-14}
\varphi_2=f_2(x)\lf(C_1+C_2\int\frac{1}{f^2_2(x)}\,dx\rg),\ee if
$\lambda_1=\frac{4}{3}\lambda_2$, where
 \[f_1(x)=\cosh^3\lf(\frac{\sqrt{a_1-a_2}}{2}\,x\rg), \quad
f_2(x)=\sinh\lf(\frac{\sqrt{a_1-a_2}}{2}\,x\rg)
\cosh^3\lf(\frac{\sqrt{a_1-a_2}}{2}\,x\rg).\]

 Thus, substituting  the functions $\varphi_1(x)$ and $\varphi_2(x)$
given  by  formulae (\ref{6-8}), (\ref{6-10}) and (\ref{6-13}) into
ansatz (\ref{6-4}), one easily obtain exact solutions of the DLV system (\ref{5-4}) (see \cite{ch-dav-2011} for details).

Let us consider an  example.

{\textbf{Example 1.}
 Using the substitution  \[u \rightarrow -bu,\ v \rightarrow -cv \ (b>0,\ c>0),\]
 one reduces the DLV system  (\ref{5-4}) to the system
  \begin{equation}\label{6-15}\begin{array}{l}
 \lbd_1 u_t = u_{xx}+u(a_1-bu-cv),\\
 \lbd_2 v_t = v_{xx}+v(a_2-bu-cv),\end{array}\end{equation}
  which describes     competition of
two species (here $a_1>0, a_2>0$ and  $\lbd_1\not=\lbd_2$).
Simultaneously this substitution transforms   solution  (\ref{6-7})
with  $C_1=0$ to the form
\begin{equation}\label{6-16}\begin{array}{l}
u(t,x)=\frac{a_1}{b}+\frac{1}{(a_1-a_2)b} \
C_2\sin\lf(\sqrt{-\beta\lambda_1}x\rg)e^{\beta t},\\
v(t,x)=\frac{1}{(a_2-a_1)c} \
C_2\sin\lf(\sqrt{-\beta\lambda_1}x\rg)e^{\beta t},
\end{array}\end{equation}  where the coefficient restrictions $\beta
\equiv \frac{a_1-a_2}{\lambda_1-\lambda_2}<0, \ a_1>0,\ a_2>0$ are
assumed. Having  this solution, we formulate the following theorem
about  the classical solution of a nonlinear BVP  involving
constant Dirichlet conditions.

\begin{figure}[h!]
\begin{center}
\includegraphics[width=7cm]{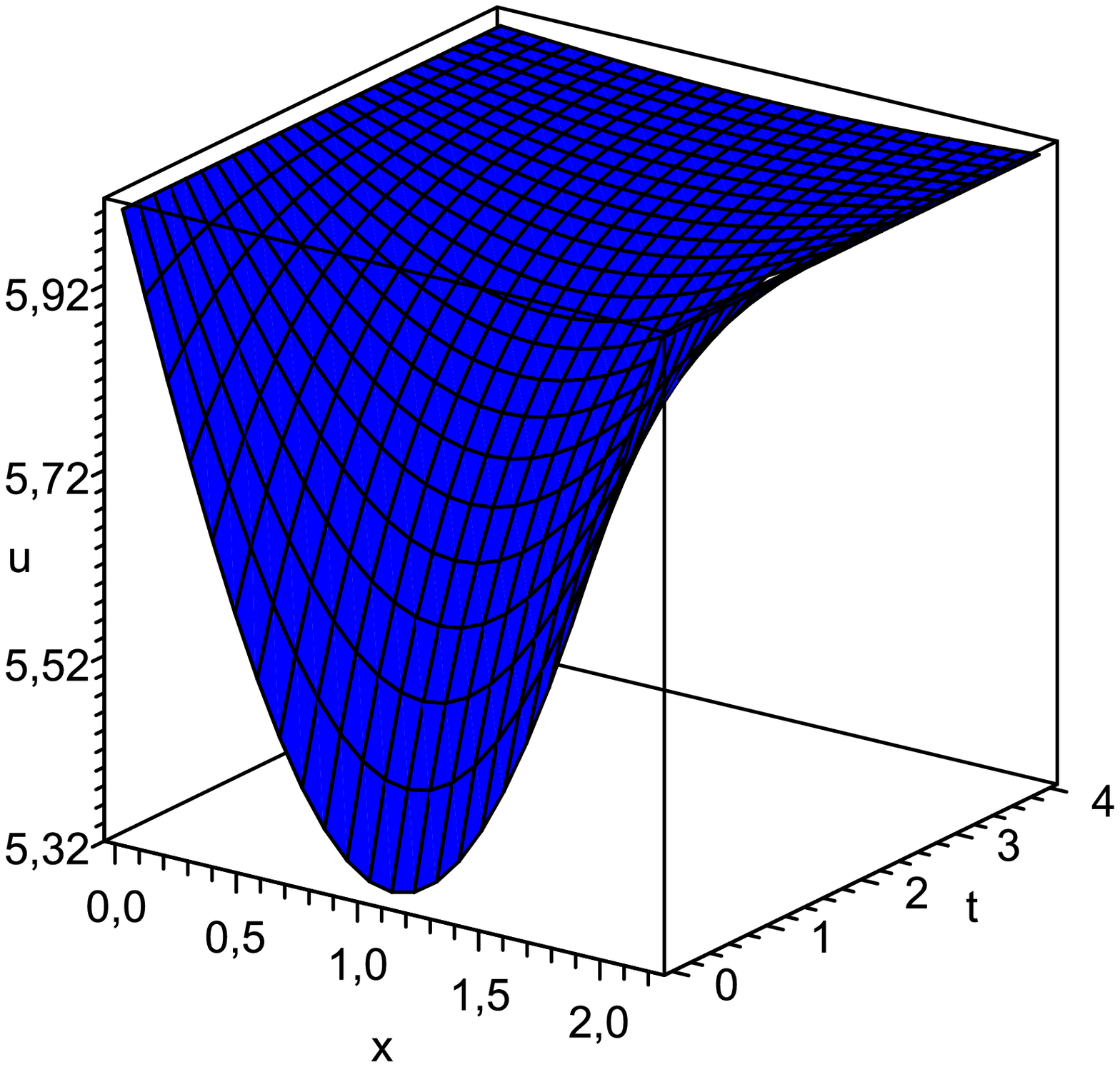}
\includegraphics[width=7cm]{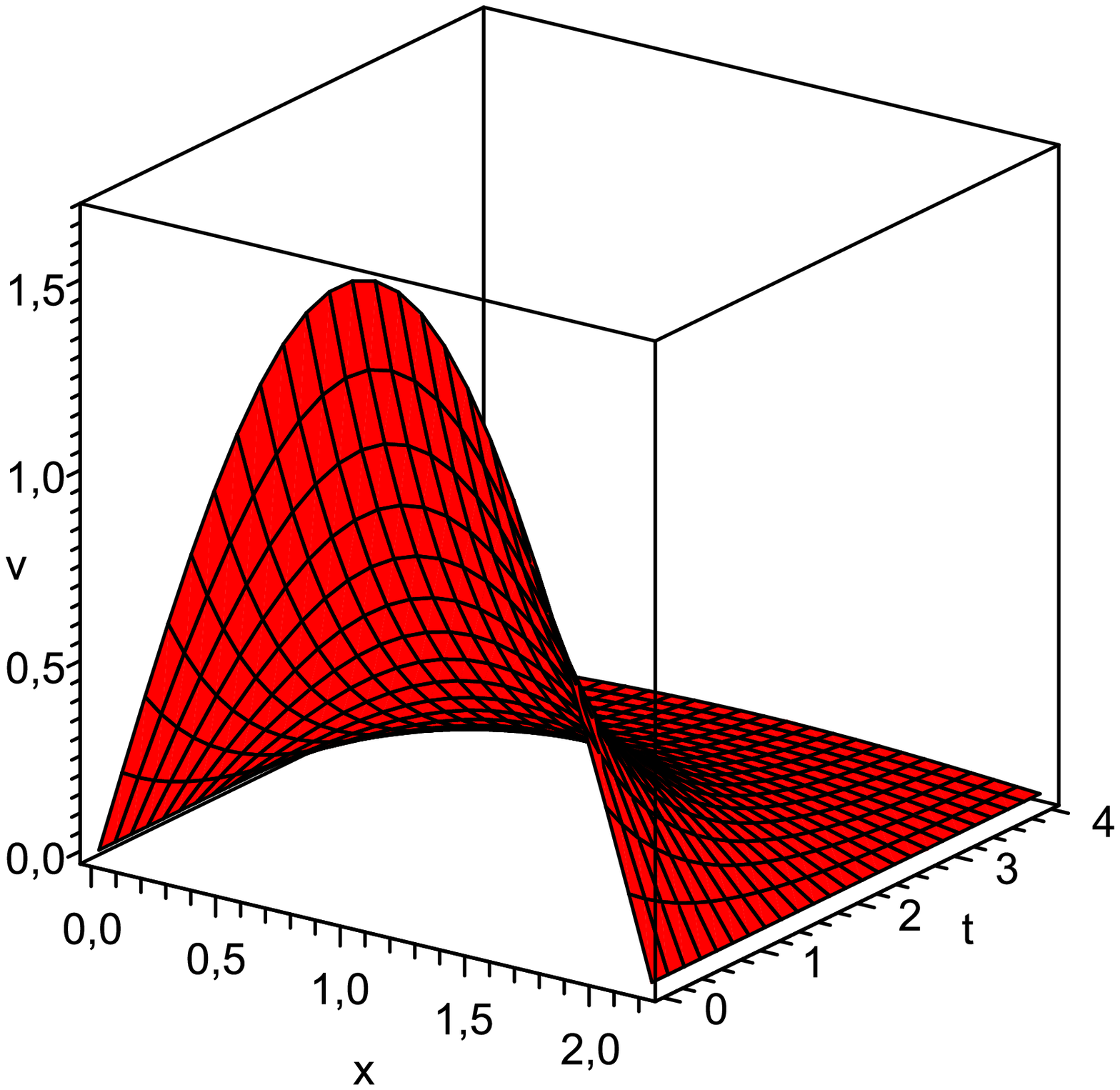}
\end{center}
\caption{Surfaces representing the  $u$ (blue) and $v$ (red)
components of solution (\ref{6-16})
 with $C_2=\frac{1}{3}, \ \beta=-1$
 of system (\ref{6-15}) with the parameters $a_1=3,\
a_2=4,\ b=\frac{1}{2}, \ c=\frac{1}{5}, \ \lambda_1=2,\ \lambda_2=1.$}
\label{6-fig1}
\end{figure}

\begin{Theorem}\cite{ch-dav-2011} The classical  solution of the nonlinear BVP  formed by the competition  system (\ref{6-15}),
 the initial profile
 \[ \ba
u(0,x)=\frac{a_1}{b}+\frac{1}{(a_1-a_2)b} \
C_2\sin\lf(\sqrt{-\beta\lambda_1}x\rg),\\
v(0,x)=\frac{1}{(a_2-a_1)c} \ C_2\sin\lf(\sqrt{-\beta\lambda_1}x\rg)
\ea\]
 and the boundary conditions
 \[\ba x=0:\  u= \frac{a_1}{b}, \ v=0,  \\
x=\frac{\pi}{\sqrt{-\beta\lambda_1}}:\  u= \frac{a_1}{b}, \ v=0
\ea\]
 in the domain  $\Omega=\lf\{ (t,x) \in (0,+ \infty )\times
\Bigl(0,\frac{\pi}{\sqrt{-\beta\lambda_1}}\Bigr)\rg\} $
 is given by formulae (\ref{6-16}).
\end{Theorem}

The solution  (\ref {6-16}) with $\beta < 0 $
 has the time asymptotic
 \[ \lf(u,\ v\rg)\rightarrow \lf(\frac{a_1}{b}, \ 0\rg), \quad
t\rightarrow +\infty.\]
 Using biological terminology, this solution  simulates    competition between  two populations
 of species  when    species $u$ eventually
dominates  while   species $v$ dies out. An example of this
competition with correctly-specified parameters  is shown in
Fig.~\ref{6-fig1}.

 Now let us consider the DLV system from Case~1 of Table~\ref{5-tab1},
  namely\,:
  \begin{equation}\label{6-17}\begin{array}{l}
\lambda_1u_t = u_{xx}+u(a_1+u+v), \\ \lambda_2v_t = v_{xx}+ v\left(a_2+\frac{\lambda_2}{\lambda_1}u+\frac{\lambda_2}{\lambda_1}v\rg), \ \lambda_1\neq \lambda_2.
 \end{array}\end{equation}
      It should be noted that  the $Q$-conditional symmetry operators $Q^u_1$ and $Q^v_1$
      of system (\ref{6-17}) lead to the same exact  solutions
      (up to discrete transformation $u\rightarrow v, \ v\rightarrow
      u$).
       Thus, we use only the operator $Q^u_1$.
The corresponding ansatz can be constructed by solving the linear
first-order PDE system
\begin{equation}\label{6-18} u_x=\frac{g^1_x}{g^1}\,u, \ v_x=-\frac{g^1_x}{g^1}\,u.\end{equation}
Integrating system (\ref{6-18}) for each form of the function $g^1$
from (\ref{5-6}), we obtain the  ansatz
\begin{equation}\label{6-19}\begin{array}{l}  u=\varphi(t)\left(\alpha_0+\alpha_1\exp\lf(-\frac{\kappa^2}{\lambda_1}\,t\rg)\sin(\kappa x)+\alpha_2\exp\lf(-\frac{\kappa^2}{\lambda_1}\,t\rg)\cos(\kappa x)\right),\\
v=\psi(t)-\varphi(t)\left(\alpha_0+\alpha_1\exp\lf(-\frac{\kappa^2}{\lambda_1}\,t\rg)\sin(\kappa x)+\alpha_2\exp\lf(-\frac{\kappa^2}{\lambda_1}\,t\rg)\cos(\kappa x)\right),
\end{array}\end{equation} if $\frac{\lambda_1a_2-\lambda_2a_1}{\lambda_1-\lambda_2}>0,$ the  ansatz
\begin{equation}\label{6-20}\begin{array}{l}  u=\varphi(t)\left(\alpha_0+\alpha_1\exp\lf(\frac{\kappa^2}{\lambda_1}\,t+\kappa x\rg)+\alpha_2\exp\lf(\frac{\kappa^2}{\lambda_1}\,t-\kappa x\rg)\right),\\
v=\psi(t)-\varphi(t)\left(\alpha_0+\alpha_1\exp\lf(\frac{\kappa^2}{\lambda_1}\,t+\kappa x\rg)+\alpha_2\exp\lf(\frac{\kappa^2}{\lambda_1}\,t-\kappa x\rg)\right),
\end{array}\end{equation} if $\frac{\lambda_1a_2-\lambda_2a_1}{\lambda_1-\lambda_2}<0,$ and the  ansatz
\begin{equation}\label{6-21}\begin{array}{l}  u=\varphi(t)\left(\alpha_0+\alpha_1x+\alpha_2x^2+2d_{1}\alpha_2t\right),\\
v=\psi(t)-\varphi(t)\left(\alpha_0+\alpha_1x+\alpha_2x^2+2d_{1}\alpha_2t\right),
\end{array}\end{equation} if $a_2=\frac{a_1\lambda_2}{\lambda_1}.$

Now three reductions of the given DLV system  to  ODE systems can be
obtained. In fact, inserting the above ans\"atze into the DLV system
(\ref{6-17}), we arrive at the ODE system
\begin{equation}\label{6-22} \lambda_1\frac{d\varphi}{dt}=\varphi\left(a_1+\psi\right) ,
 \ \lambda_2\frac{d\psi}{dt}=\left(a_2+\frac{\lambda_2}{\lambda_1}\,\psi\right)\psi+\alpha_0\lf(\frac{a_1\lambda_2}{\lambda_1}-a_2\rg)\,\varphi,
\end{equation} in the cases of formulae (\ref{6-19}) and (\ref{6-20}),
while the system
\begin{equation}\label{6-23} \lambda_1\frac{d\varphi}{dt}=\varphi\left(a_1+\psi\right) , \ \lambda_2\frac{d\psi}{dt}=\frac{\lambda_2}{\lambda_1}\left(a_1+\psi\right)\psi-2
\alpha_2\lf(\lambda_1-\lambda_2\rg)\,\varphi,
\end{equation} is obtained in the case of (\ref{6-21}).
Here $\varphi(t)$ and $\psi(t)$ are to-be-determined functions.

It was proved that each of the ODE systems  (\ref{6-22}) and
(\ref{6-23}) can be integrated by reducing to a single second-order
ODE (see \cite{ch-dav-2021}
 for details). Here we present  exact solutions of the DLV system  (\ref{6-17})
 with $a_1a_2\neq0$ and  $\frac{\lambda_1a_2-\lambda_2a_1}{\lambda_1-\lambda_2}>0$,
 namely: \begin{equation}\label{6-24}\begin{array}{l}
u(t,x)=\frac{a_1\exp\lf(\frac{a_1}{\lambda_1}\,t\rg)}{C_1-\alpha_0\exp\lf(\frac{a_1}{\lambda_1}\,t\rg)+
C_2\lambda_2\exp\lf(\frac{a_2}{\lambda_2}\,t\rg)}
\left(\alpha_0+\alpha_1\exp\lf(-\frac{\kappa^2}{\lambda_1}\,t\rg)\sin(\kappa x)+\alpha_2\exp\lf(-\frac{\kappa^2}{\lambda_1}\,t\rg)\cos(\kappa x)\right), \medskip  \\
v(t,x)=\frac{\alpha_0a_1\exp\lf(\frac{a_1}{\lambda_1}\,t\rg)-C_2a_2\lambda_1\exp\lf(\frac{a_2}{\lambda_2}\,t\rg)}{C_1-
\alpha_0\exp\lf(\frac{a_1}{\lambda_1}\,t\rg)+
C_2\lambda_2\exp\lf(\frac{a_2}{\lambda_2}\,t\rg)}-u(t,x).
\end{array}\end{equation} Here $\alpha_i$, $C_1$ and $C_2$ are arbitrary
constants, which should be specified using additional
conditions/requirements  satisfied  by the exact solution
(\ref{6-24}).

{\textbf{Example 2.} Using the transformation  $u\rightarrow-bu,\
v\rightarrow-cv$ and introducing the notation
$\alpha_0\rightarrow-b\,\alpha_0,\ \alpha_1\rightarrow-b\,\alpha_1,$
 one reduces the  DLV system  (\ref{6-17}) to the form
  \begin{equation}\label{6-25}
  \begin{array}{l}
\lambda_1u_t = u_{xx}+u(a_1-b\,u-c\,v), \\ \lambda_2v_t = v_{xx}+ v\left(a_2-\frac{\lambda_2b}{\lambda_1}u-\frac{\lambda_2c}{\lambda_1}v\rg).
\end{array}\end{equation}
The nonlinear system  (\ref{6-25}) with positive parameters $a_1, \ a_2,
 \ b$ and $c$  can be applied  for modeling   competition of two
 population of species.
Solution  (\ref{6-24}) (we set $\alpha_2=0$  just for simplicity)
after the above transformation reads as follows
\begin{equation}\label{6-26}\begin{array}{l}
u(t,x)=\frac{a_1\exp\lf(\frac{a_1}{\lambda_1}\,t\rg)}{C_1+\alpha_0b\exp\lf(\frac{a_1}{\lambda_1}\,t\rg)+
C_2\lambda_2\exp\lf(\frac{a_2}{\lambda_2}\,t\rg)}
\left(\alpha_0+\alpha_1\exp\lf(\frac{\lambda_2a_1-\lambda_1a_2}{\lambda_1(\lambda_1-\lambda_2)}\,t\rg)
\sin\lf(\sqrt{\frac{\lambda_1a_2-\lambda_2a_1}{\lambda_1-\lambda_2}}\,x\rg)\right), \medskip  \\
v(t,x)=\frac{1}{c}\frac{\alpha_0a_1b\exp\lf(\frac{a_1}{\lambda_1}\,t\rg)+C_2a_2\lambda_1\exp\lf(\frac{a_2}{\lambda_2}\,t\rg)}{C_1+
\alpha_0b\exp\lf(\frac{a_1}{\lambda_1}\,t\rg)+
C_2\lambda_2\exp\lf(\frac{a_2}{\lambda_2}\,t\rg)}-\frac{b}{c}\,u(t,x).
\end{array}\end{equation}

In order to provide a biological interpretation, we introduce the
following requirements: the  $u$ and $v$ components are bounded and
nonnegative in a domain because they represent densities of species.
Let us consider the domain $\Omega=\left\{ (t,x) \in (0,+ \infty
)\times (-\infty,+ \infty)\right\}$. It  can be shown that both
components are  bounded and nonnegative if the coefficient
restrictions
 \[\alpha_0>\left|\alpha_1\right|, \ C_2>\max\left\{-\frac{\alpha_0b+C_1}{\lambda_2}, \ \frac{ba_1\left|\alpha_1\right|}{a_2\lambda_1}\right\}\]  hold.
We also note that the exact  solution (\ref{6-26})
 possesses the  asymptotical behavior
  \begin{equation}\label{6-27}\begin{array}{l}   \begin{array}{l} (u,\,v)  \rightarrow  \left(\frac{a_1}{b},\,0\right), \ \texttt{if} \ a_1\lambda_2>a_2\lambda_1,\\
(u,\,v)  \rightarrow  \left(0,\,\frac{a_2\lambda_1}{c\lambda_2}\right), \ \texttt{if} \  a_1\lambda_2<a_2\lambda_1,
\end{array} \ \texttt{as} \  t  \rightarrow +\infty.
\end{array}  \end{equation}
Now one realizes that $\left(\frac{a_1}{b},\,0\right)$  and
$\left(0,\,\frac{a_2\lambda_1}{c\lambda_2}\right)$ are steady state points
 of the competition model (\ref{6-25})  and the  asymptotical
 behavior (\ref{6-27}) is in agreement with the qualitative theory
 of this model (see, e.g., \cite{lam-20} and papers cited therein).

\begin{figure}[h!]
\begin{center}
\includegraphics[width=7cm]{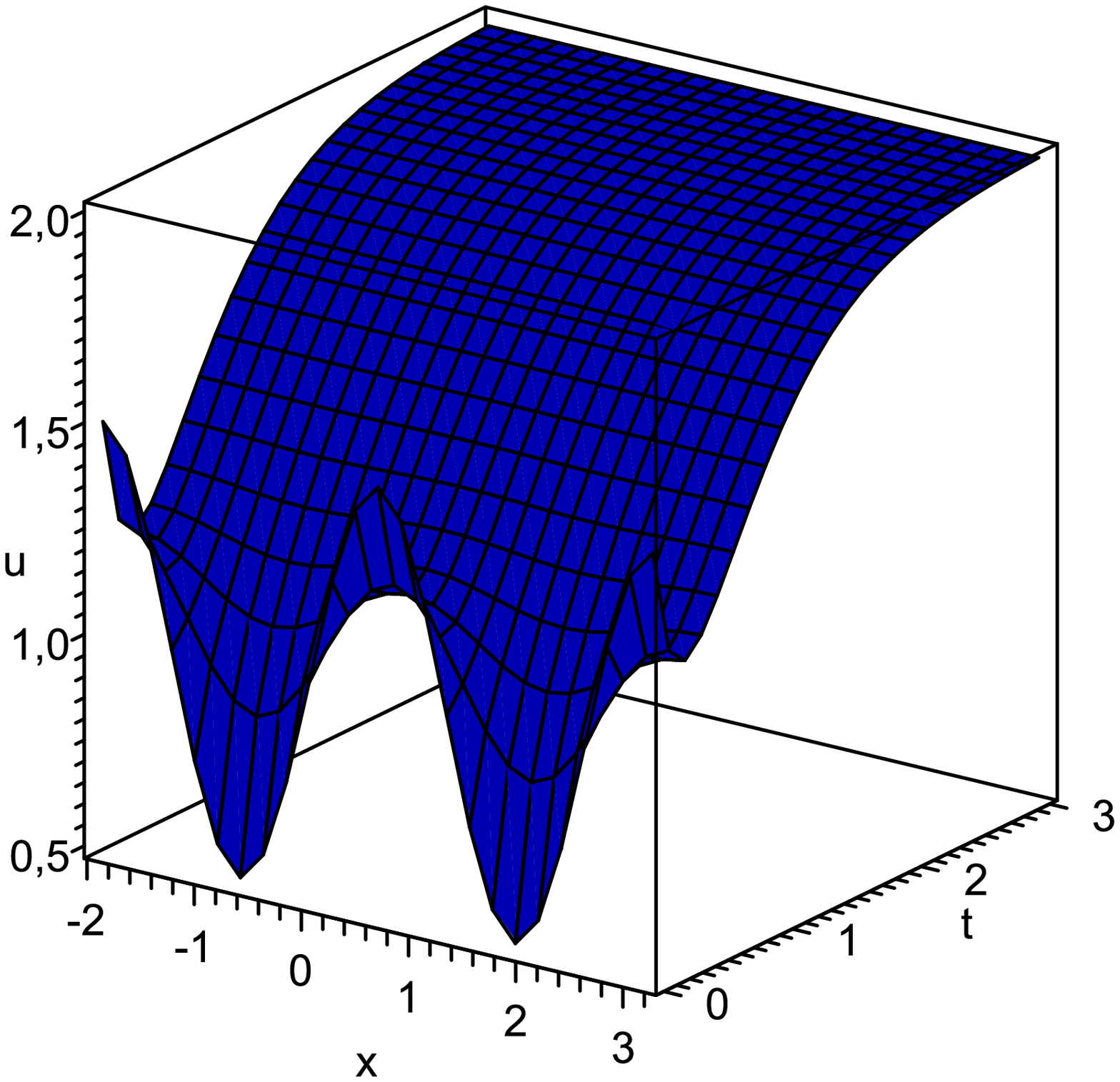}
\includegraphics[width=7cm]{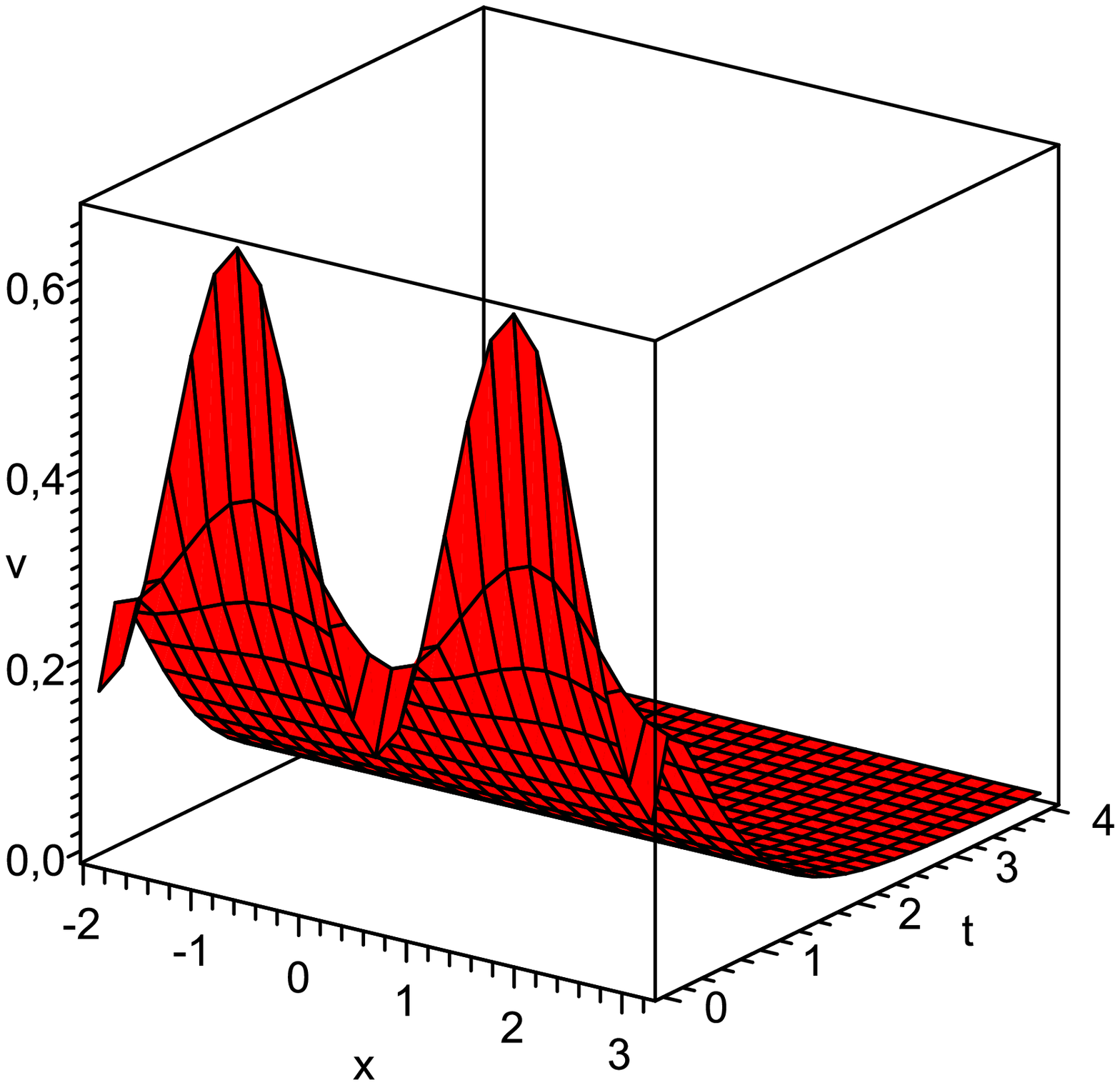}
\end{center}
\caption{Surfaces representing the  $u$ (blue) and $v$ (red)
components of solution (\ref{6-26})
 with $C_1=-2, \ C_2=5, \ \alpha_0=2, \ \alpha_1=1$
 of system (\ref{6-25}) with the parameters $a_1=3,\
a_2=2,\ b=\frac{3}{2}, \ c=3, \ \lambda_1=\frac{3}{4},\ \lambda_2=1.$}
\label{6-fig2}
\end{figure}

In real-world applications,  competition usually occurs  in bounded
domains. Let us consider the domain $\Omega_*=\left\{ (t,x) \in (0,+
\infty )\times (A,B)\right\},  \  -\infty<A<B<+\infty$. Typically,
zero flux conditions are prescribed  at the boundaries:
\[\begin{array}{l}
x=A: \, u_x=0, \, v_x=0, \\
x=B: \, u_x=0, \, v_x=0. \end{array} \] The zero flux conditions
reflect a natural assumption that the competing species cannot cross
the boundaries (e.g., a wide river  could be a natural obstacle).
One easily checks that the exact solution (\ref{6-26})
 satisfies the boundary conditions provided
\[ A=\frac{\pi}{\kappa}\left(\frac12+m_1\right), \,
B=\frac{\pi}{\kappa}\left(\frac12 +m_2\right), \ m_1<m_2. \] Here
$m_1$ and $m_2$ are arbitrary integer parameters and
$\kappa=\sqrt{\frac{\lambda_1a_2-\lambda_2a_1}{\lambda_1-\lambda_2}}.$
 Thus, we conclude that  the exact solution (\ref{6-26})
 with correctly-specified parameters simulates
  the competition of two population of species in the bounded domain.
   An example is presented in Fig.~\ref{6-fig2}.

\section{\bf Exact solutions of the three-component DLV system}\label{sec-7}

This section is a natural continuation of the previous one. The only
difference is that here a three-component DLV system is studied
instead of a two-component system. It should be pointed out that the
three-component DLV system (\ref{3-3}) admits a much wider set of
$Q$-conditional symmetries
compared to  the two-component analogue. One may  apply each
$Q$-conditional symmetry arising  in Table~\ref{5-tab2} in order to
find exact solutions for the biologically motivated DLV system.

One notes that the DLV systems arising in  Cases 1--4  of
Table~\ref{5-tab2}  can be reduced to those modeling different types
of interaction between three populations of species (cells,
chemicals etc.). Here we examine in details only Case 4 because the
corresponding symmetry operators have the most complicated structure
(Cases 1--3 can be examined in a quite similar way) and present the
results derived in \cite{ch-dav2013}.
 Obviously, the system from Case~4 of
 is   reducible by  the substitution  $u \to -bu, \ v \to -cv, \
w \to -ew$   to the system
\be\label{7-1}\ba \lbd_1 u_t =  u_{xx}+u(a_1-bu-cv-ew),\\
\lbd_2 v_t = v_{xx}+ v(a_2-bu-cv-ew),\\ \lbd_3 w_t =
w_{xx}+w(a_3-bu-cv-ew), \ea\ee where the parameters $a_k, \ b, \ c$
and $e$ are  positive constants. System (\ref{7-1}) can be used, in
particular,  for modeling  three competing  species in the
population dynamics.

Let as assume that the coefficients $a_k$ and $\lbd_k \ (k=1,2,3)$
satisfy the restrictions presented in Case~4 of Table \ref{5-tab2}.
It means that  the system admits the symmetry  operators $Q^4_i \
(i=1,\dots, 6)$, which have the same structure. Substituting $u \to
-bu, \ v \to -cv, \ w \to -ew$ into, e.g.,  the $Q$-conditional
symmetry operator $Q^4_1$ we obtain
 \be\label{7-2}Q^4_1  \to
Q=\p_t+\frac{a_1-a_2}{\lambda_1-\lambda_2}\,u\lf(\p_u-\frac{b}{c}\,\p_v\rg)+\alpha
b\, u\lf(\frac{1}{c}\,\p_v-\frac{1}{e}\,\p_w\rg). \ee
 So, using the standard algorithm to reduce the given PDE system  to
 an ODE
  system  via  the known operator  (\ref{7-2}),
 one can easily obtain
 the ansatz
\be\label{7-3}\ba u=\frac{\varphi_1(x)}{b}\,e^{\delta t},\medskip \\ v=\frac{\varphi_2(x)}{c}+
\lf(\frac{\alpha}{\delta}-1\rg)\frac{\varphi_1(x)}{c}\, e^{\delta t},\medskip
\\ w=\frac{\varphi_3(x)}{e}-\frac{\alpha}{e\delta}\,\varphi_1(x)\,e^{\delta t},
\ \, \delta=\frac{a_1-a_2}{\lambda_1-\lambda_2}\not=0,\ea\ee where
$\varphi_1(x)$, $\varphi_2(x)$ and $\varphi_3(x)$  are
to-be-determined functions. Substituting ansatz (\ref{7-3}) into
(\ref{7-1}) and taking into account the restriction
$$(\lambda_2-\lambda_3)a_1-(\lambda_1-\lambda_3)a_2+(\lambda_1-\lambda_2)a_3=0$$
(see Case 4 of Table~\ref{5-tab2}), we arrive at  the reduced system
of ODEs
\be\label{7-4}\ba\varphi_1''+\varphi_1\lf(\frac{\lambda_1a_2-\lambda_2a_1}{\lambda_1-\lambda_2}-\varphi_2-\varphi_3\rg)=0,
 \\
\varphi_2''+\varphi_2\lf(a_2-\varphi_2-\varphi_3\rg)=0,
\\
\varphi_3''+\varphi_3\lf(a_3-\varphi_2-\varphi_3\rg)=0.\ea\ee Thus,
exact solutions of the three-component competition system
(\ref{7-1}) can be obtained  by  substitution  of arbitrary
solutions of system (\ref{7-4}) into ansatz (\ref{7-3}).

System (\ref{7-4}) is three-component system of  nonlinear
second-order ODEs. To the best of our knowledge,  its  general
solution is unknown. Let us assume that the triplet
$(\varphi^0_1(x), \varphi^0_2(x), \varphi^0_3(x))$ is a particular
solution of (\ref{7-4}). Moreover,  we assume that  the functions
$\varphi^0_k$ are nonnegative and bounded on a space interval $I$.
Having this, we  observe that the exact solution (\ref{7-3}) with
$\varphi_k= \varphi^0_k \ (k=1,2,3)$ tends to the steady-state
solution $\lf(0, \frac{\varphi^0_2}{c}, \frac{\varphi^0_3}{e}\rg)$
of the DLV system (\ref{7-1}) with $\delta<0$ provided $t \to
+\infty $. In the general case, the solution $\lf(0,
\frac{\varphi^0_2}{c}, \frac{\varphi^0_3}{e}\rg)$ produces a curve
in the phase space $(u, v, w)$, which lies in the plane $(0, v, w)$.
So,   considering  the competition of three populations  at the
space interval $I$, we  conclude that the exact  solution
(\ref{7-3}) with $\varphi_k= \varphi^0_k$ $(k=1,2,3)$ describes such
competition when species $u$ dies out while species $v$ and $w$
 coexist. In particular,  a limit cycle  may occur
 if the concentrations  $v= \frac{\varphi^0_2(x)}{c}$
  and $w= \frac{\varphi^0_3(x)}{e}$    form  a closed  curve.

Let us consider an example in  the  case  when $\varphi^0_2(x)$ and
$\varphi^0_3(x)$ are constants. It can  easily checked that the
constant solution $\varphi_2=v_0, \ \varphi_3=a_2-v_0$ of the second
and third equations of (\ref{7-4}) with $a_2=a_3$, generates the
following solution of the three-component competition system
(\ref{7-1}) with $ a_1\neq a_2=a_3$:
\be\label{7-5}
\ba u=\frac{\varphi_1(x)}{b}e^{\delta t},
\\ v=\frac{v_0}{c}+ \frac{1}{c}
\lf(\frac{\alpha}{\delta}-1\rg)\varphi_1(x)e^{\delta t},
\\ w=\frac{a_2-v_0}{e} -\frac{\alpha\varphi_1(x)}{e\delta}e^{\delta
t}, \ea\ee where  $\varphi_1(x)$ is a solution of the linear
ODE
 \be\label{7-6} \varphi_1''-
 \lambda_2\delta\,\varphi_1=0. \ee

Interestingly, the exact  solution  (\ref{7-5}) is not obtainable by
any Lie symmetry because system (\ref{7-1})
  admits    the  Li  algebra  (\ref{3-2}),
  so that only  traveling wave solutions can be constructed.

  We point out that  the general solution of ODE (\ref{7-6})
  essentially depends on the sign of $\delta$. In the case $\delta>0$,
    unbounded (in time)  solutions (see formulae (\ref{7-5}))
   are obtained  and it is unlikely that  they can describe
   a realistic competition
   between three populations.

On the other hand, equation  (\ref{7-6}) with   $\delta<0$  has the
general solution
\be\label{7-7}\varphi(x)=C_1\cos\lf(\sqrt{-\delta\lambda_2}\,x\rg)
+C_2\sin\lf(\sqrt{-\delta\lambda_2}\,x\rg), \ee where $C_1$ and
$C_2$ are arbitrary constants. Setting, for example,
 $C_1=0$  and  $C_2=1$  in  (\ref{7-7})  and substituting
  $\varphi(x)$ into  (\ref{7-5}), we obtain the exact  solution
\be\label{7-8}\ba u=\frac{1}{b}\sin\lf(\sqrt{-\delta\lambda_2}\,x\rg)e^{\delta
t}, \\ v=\frac{v_0}{c}+ \frac{1}{c}
\lf(\frac{\alpha}{\delta}-1\rg)\sin\lf(\sqrt{-\delta\lambda_2}\,x\rg)e^{\delta
t},  \\ w=\frac{a_2-v_0}{e}
-\frac{\alpha}{e\delta}\sin\lf(\sqrt{-\delta\lambda_2}\,x\rg)e^{\delta
t} \ea\ee of  system (\ref{7-1}) with  $ a_1\neq a_2=a_3, \
\delta=\frac{a_1-a_2}{\lambda_1-\lambda_2}.$

Let us provide a biological interpretation of the exact  solution
(\ref{7-8}). For these purposes, we assume that the competition
between three populations occurs at the space interval $I=\lf[0,
\frac{\pi}{\sqrt{-\delta\lambda_2}}\rg]$.  Obviously, the components of the
exact solution  (\ref{7-8}) satisfy the boundary conditions
 \[\ba x=0:  \ u= 0, \ v=\frac{v_0}{c}, \
w=\frac{a_2-v_0}{e}; \\
  x= \frac{\pi}{\sqrt{-\delta\lambda_2}}: \ u= 0, \ v=\frac{v_0}{c}, \ w=\frac{a_2-v_0}{e}.  \ea\]
These conditions predict that the densities  of  the species $u, \
v$ and $w$ are constant values at the boundaries (it means that an
artificial
 regulation of the population densities holds in a vicinity of the
 $x=0$ and $x= \frac{\pi}{\sqrt{-\delta\lambda_2}}$ points).
  Moreover, this
exact  solution tends to the steady-state
point $\lf(0, \ \frac{v_0}{c},\ \frac{a_2-v_0}{e}\rg)$  if
$t\rightarrow +\infty.$

It can be checked that all the components in (\ref{7-8}) are bounded
and nonnegative for an arbitrary given  $t>0$ and $x \in I$ provided
the additional  restrictions \[\ba
  0\leq v_0\leq a_2-\frac{\alpha}{\delta}, \  \mbox{if} \  \alpha\leq\delta,
  \\
1-\frac{\alpha}{\delta}\leq v_0\leq a_2-\frac{\alpha}{\delta}, \
\mbox{if} \ \delta<\alpha\leq0, \\
 1-\frac{\alpha}{\delta}\leq v_0\leq a_2,  \   \mbox{if} \
\alpha>0  \ea\]
 hold.
Thus, the exact  solution  (\ref{7-8})   describes  the following
scenarios of the competition between  three  species:
\begin{itemize}
\item[\emph{\textbf{(i)}}] \hskip0.2cm  species $v$ and $w$ eventually coexist
while  species $u$ dies out provided \[0<v_0<a_2;\]
\item[\emph{\textbf{(ii)}}] \hskip0.2cm  species $v$  eventually dominates while
 species $u$ and $w$ die out provided \[v_0=a_2;\]
\item[\emph{\textbf{(iii)}}] \hskip0.2cm  species $w$ eventually dominates while
 species $u$ and $v$ die out provided \[v_0=0.\]
\end{itemize}
 Examples of scenarios
$\emph{\textbf{(i)}}$  and $\emph{\textbf{(ii)}}$ are presented in
Fig.~\ref{7-fig1} and Fig.~\ref{7-fig2}, respectively.

\begin{figure}[h!]
\begin{center}
\includegraphics[width=7cm]{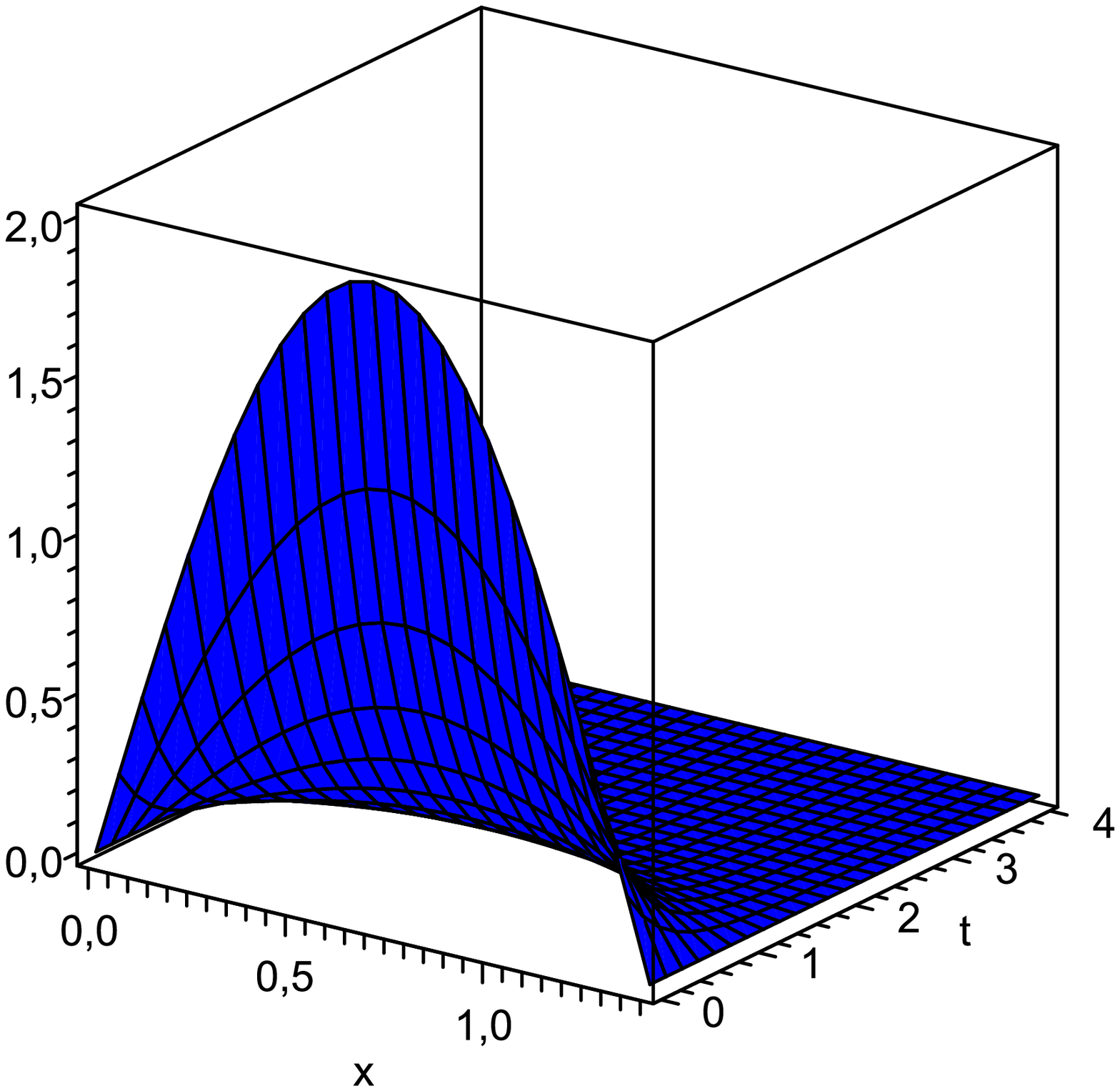}
\includegraphics[width=7cm]{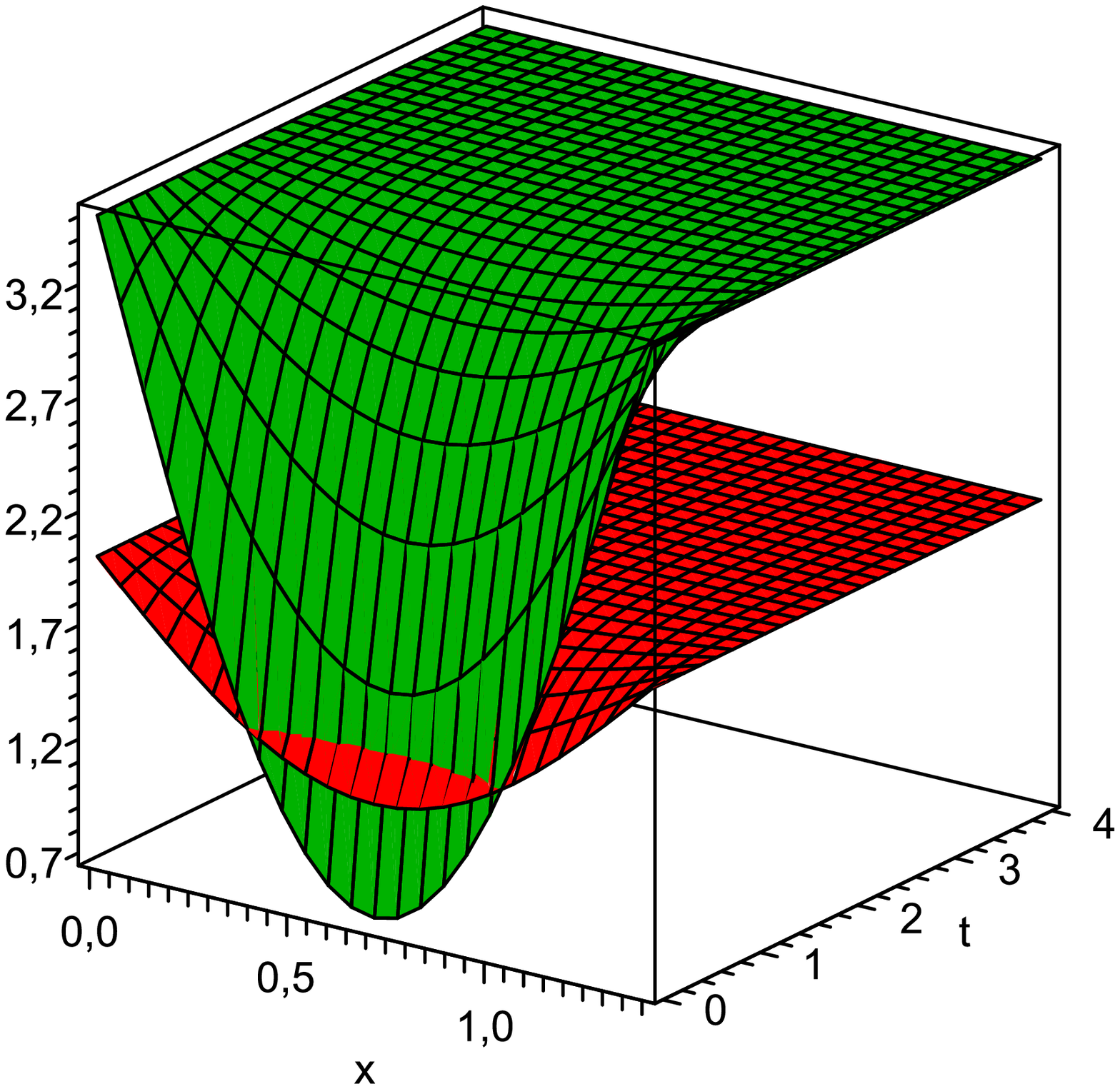}
\end{center}
\caption{Surfaces representing the $u$ (blue), $v$ (red) and $w$
(green) components  of solution (\ref{7-8})  with $\alpha=-1, \
v_0=\frac{3}{2}, \ \delta=-\frac{5}{2}$
 of system (\ref{7-1}) with the parameters $a_1=\frac{9}{2},\
a_2=a_3=2,\ b=\frac{1}{2}, \ c=\frac{3}{4}, \ e=\frac{1}{7}, \ \lambda_1=1,\ \lambda_2=\lambda_3=2.$}
\label{7-fig1}
\end{figure}

\begin{figure}[h!]
\begin{center}
\includegraphics[width=7cm]{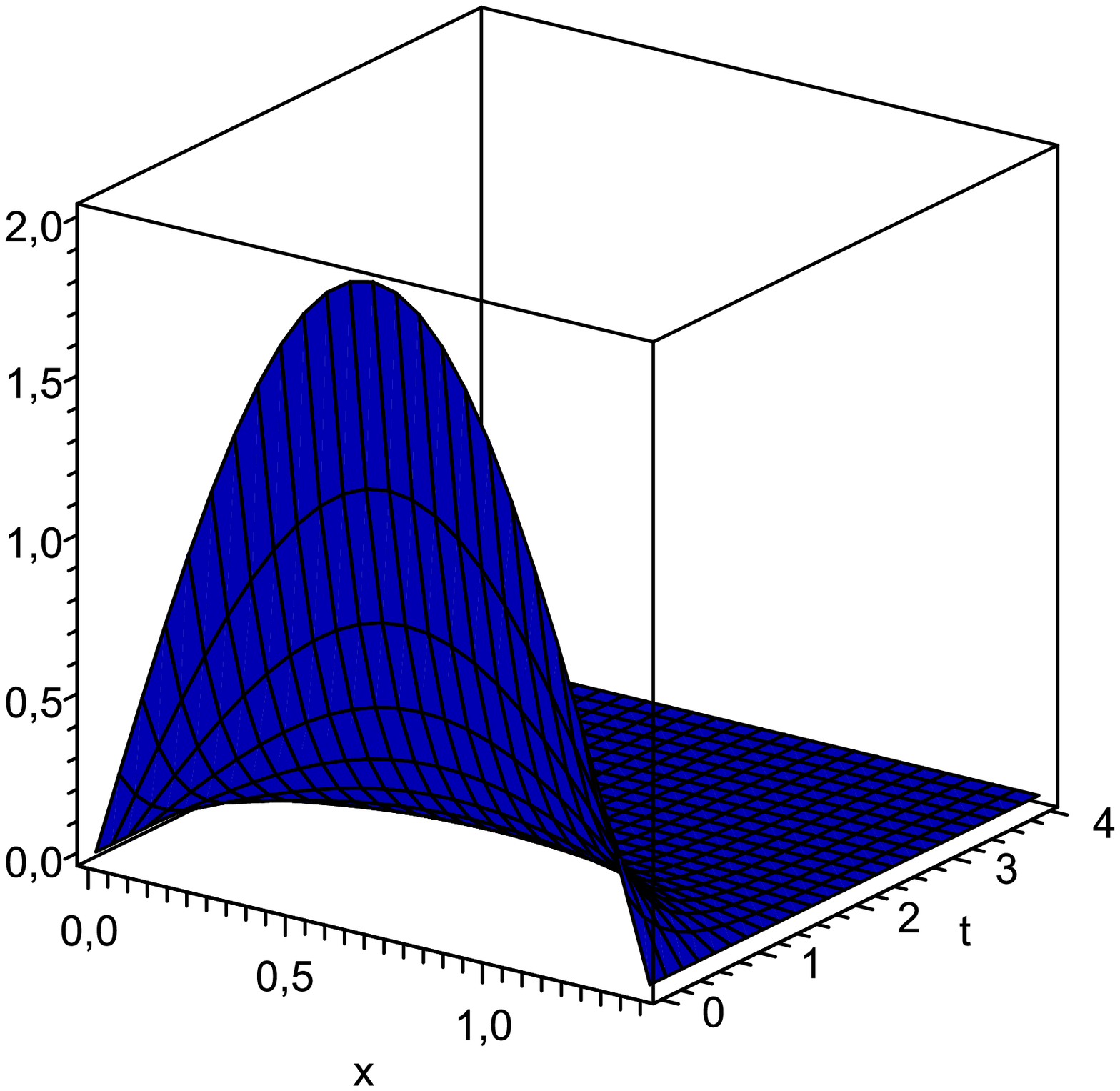}
\includegraphics[width=7cm]{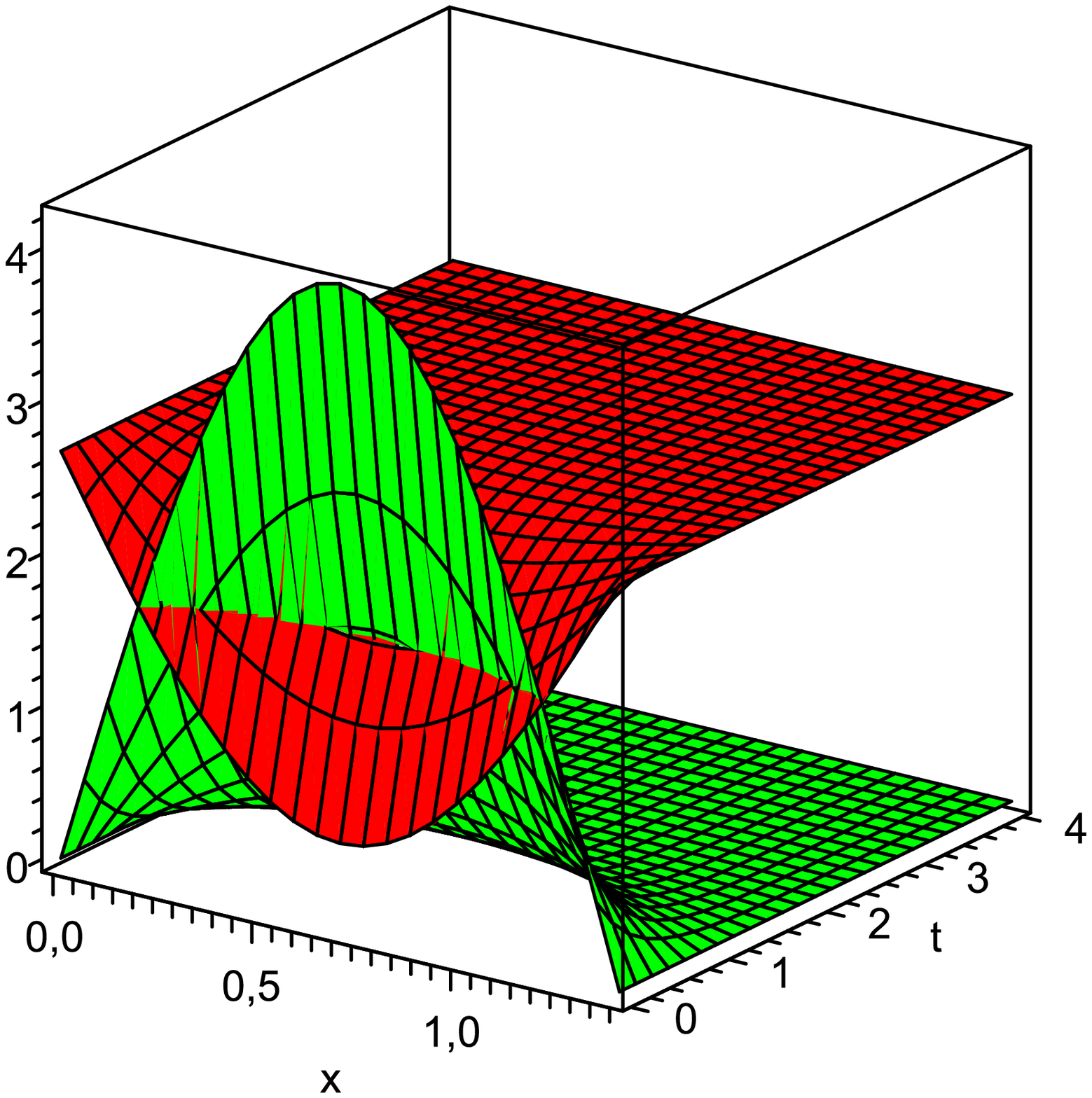}
\end{center}
\caption{Surfaces representing the  $u$ (blue), $v$ (red) and $w$
(green) components of solution (\ref{7-8})
 with $\alpha=\frac{3}{2}, \ v_0=2, \ \delta=-\frac{5}{2}$
 of system (\ref{7-1}) with the parameters $a_1=\frac{9}{2},\
a_2=a_3=2,\ b=\frac{1}{2}, \ c=\frac{3}{4}, \ e=\frac{1}{7}, \ \lambda_1=1,\ \lambda_2=\lambda_3=2.$}
\label{7-fig2}
\end{figure}

\medskip

An essential progress in constructing exact solutions of the
three-component
 DLV system  was achieved in  \cite{hung16}. New exact solutions  were discovered
 when system (\ref{3-3}) involves  equal diffusivities (i.e. $
 \lbd_1=\lbd_2=\lbd_3$), positive $a_i$ and  negative
 $b_i, \ c_i,$ and $ e_i$ parameters  (i.e. describes competition of three
 populations). In this case, the DLV system (\ref{3-3}) is reducible to the form
\be\label{7-9}\ba u_t =  u_{xx}+u(1-u-c_1v-e_1w),\\
v_t = v_{xx}+ c_2v(1-b_2u-v-e_2w),\\ w_t =
w_{xx}+e_3w(1-b_3u-c_3v-w) \ea\ee Assuming that linear terms
$1-u-c_1v-e_1w$, $1-b_2u-v-e_2w$  and  $1-b_3u-c_3v-w$ arising in
the RHS of the system  are linearly dependent,  the following family
of exact solutions was derived
  \be\label{7-10}\ba
u(t,x)=\frac{c_1-1}{c_1b_2-1}+\frac{e_1-c_1e_2}{c_1b_2-1}\lf(w_0+\frac{1}{\sqrt{4\pi
t}}\int\limits_{ -\infty}^{\ \ \infty}\exp\lf(-\frac{(x-y)^2}{4t}
\rg)f(y)dy\rg),\medskip\\
v(t,x)=\frac{b_2-1}{c_1b_2-1}+\frac{e_2-b_2e_1}{c_1b_2-1}\lf(w_0+\frac{1}{\sqrt{4\pi t}}\int\limits_{
-\infty}^{\ \ \infty}\exp\lf(-\frac{(x-y)^2}{4t}
\rg)f(y)dy\rg),\medskip\\
w(t,x)=w_0+\frac{1}{\sqrt{4\pi t}}\int\limits_{ -\infty}^{\ \
\infty}\exp\lf(-\frac{(x-y)^2}{4t} \rg)f(y)dy, \ea\ee where $w_0$ is
an arbitrary constant, while $f(y)$ is an arbitrary  continuous
function such that the integral in the RHS of (\ref{7-10})
converges.

Although this result is formulated in the form of a cumbersome
theorem (see Theorem 2.1 in \cite{hung16}), the main idea is very
simple and was implicitly used earlier in \cite{ch-du-04}. In fact,
according to the assumption, there exist  constants $A, \ B,$ and
$C$ such that
\[  A(1-u-c_1v-e_1w)+B(1-b_2u-v-e_2w) + C(1-b_3u-c_3v-w)=0.
 \]
 So, taking the  linear combination of equations from  (\ref{7-10}),
 we exactly arrive at the linear diffusion equation
\be\label{7-10*}  U_t=U_{xx}, \quad  U=Au+Bv+Cw \ee with
correctly-specified $A, \ B$ and $C$.  Obviously, the integral in
the RHS of (\ref{7-10}) is the well-known solution of (\ref{7-10*}).

In particular case, solution (\ref{7-10}) with
$f(y)=\beta\sin(\gamma y)$ (here $\beta$ and $\gamma$ are nonzero
constants) takes the form \cite{hung16}
\be\label{7-11}\ba u(t,x)=\frac{c_1-1}{c_1b_2-1}+\frac{e_1-c_1e_2}{c_1b_2-1}
\lf(w_0+\beta\sin(\gamma x)e^{-\gamma^2t}\rg),\medskip\\
v(t,x)=\frac{b_2-1}{c_1b_2-1}+\frac{e_2-b_2e_1}{c_1b_2-1}\lf(w_0+\beta\sin(\gamma x)e^{-\gamma^2t}\rg),\medskip\\
w(t,x)=w_0+\beta\sin(\gamma x)e^{-\gamma^2t}. \ea\ee

It can be  seen  that the exact solution (\ref{7-11}) is a
generalization of solution  (\ref{7-8}) on the case when all the
diffusivities are equal.


\section{\bf Conclusions}\label{sec-8}

This work summarizes all  known results (up to this date)  about
methods of  integration of the classical Lotka--Volterra systems
with diffusion and presents a wide range  of exact solutions, which
are the most important from applicability point of view. To the best
of our knowledge, it is the first attempt in this direction. Because
the DLV systems are used for mathematical modeling of an enormous
variety of processes in ecology, biology, medicine, chemistry, etc.
(see, e.g.,  well-known books \cite{britton, mur2, mur2003,
okubo,ku-na-ei-16, fife-79, aris-75I}), we believe that it is an
appropriate time for such  kind of a review.

We would like to point out  that exact solutions always play an
important role for any nonlinear model describing real-world
processes.
At the present time, there is no general theory for integrating
nonlinear PDEs (system of PDEs). Thus,  construction of particular
exact solutions for these equations is a highly  nontrivial and
important problem. Identifying  exact solutions in a closed  form
that have a physical (chemical, medical, biological etc.)
interpretation is of fundamental importance. Even exact solution
with questionable applications can be important for proper
examination of software packages devoted to numerical solving of
systems of PDEs. The obtained exact solutions can also be used as
test problems to estimate the accuracy of approximate analytical
methods for solving of  boundary value problems for PDEs.

In this review, the main attention was paid to symmetry-based
methods for exact solving the classical Lotka--Volterra systems with
diffusion. We briefly presented the relevant theory
(Section~\ref{sec-2}) and application of the theory to find Lie
symmetries  of the two- and three-component LV systems
(Section~\ref{sec-3}). Furthermore, we applied the simplest Lie
symmetries for constructing plane wave solutions, especially
traveling fronts, which are the most popular type of exact solutions
in the case of nonlinear evolution equations (Section~\ref{sec-4}).
We also presented the most interesting traveling waves derived by
other authors, including those from the pioneering work
\cite{rod-mimura-2000}. It turns out that  Lie symmetries have
rather a limited efficiency if one looks for exact solutions of the
DLV systems, therefore we  derived  wide families  of conditional
symmetries of the DLV systems under study (Section~\ref{sec-5}).
Finally, the conditional symmetries obtained were used to construct
exact solutions with more complicated structures than the traveling
 fronts. Moreover, examples of  applications of  some exact solutions
 for solving real-world models based on the DLV systems are successfully
 demonstrated  (Sections~\ref{sec-6} and~\ref{sec-7}). We also presented an interesting
 family of exact solutions derived in  \cite{hung16} by an ad hoc
 technique,  which seems to be not related with symmetry-based
 methods.

In conclusion, we would like to highlight some unsolved problems. In
this review, a majority of exact solutions are related to the DLV
systems describing the competition of two (three) populations of
species (cells). However, there are other types of interaction
between species, cells, chemicals etc. In particular, the nonlinear
system (\ref{1-2}), in  which all the   parameters $a_i$ and
$b_{ij}$ are nonnegative,  is a model describing mutualism or
cooperation (see, e.g., \cite{britton, ugalde-2021}). Obviously, the
solutions presented  in this work are  useful for interactions of
such type
 as well. On the other hand, these solutions are not
applicable for the third most common type of interaction between
species (cells, chemicals, etc.) leading to  prey-predator models.
In the two-component prey-predator model, the parameters satisfy the
following  typical  restrictions $a_1a_2<0, \ c_1b_2<0$, \
$b_1\leq0$
 (see the DLV system
(\ref{3-1})). It can be seen that Tables~\ref{3-tab1}, \ref{5-tab3}
 and \ref{5-tab1} do not contain such types of systems, therefore the
relevant exact solutions cannot be found. Moreover, we have checked
that the exact solutions derived in  the following studies
\cite{rod-mimura-2000,ch-du-04,hung12, kudrya-15, hung11,
chen-hung12, ch-dav-2011,ch-dav2013}
cannot describe   the prey-predator interaction either
 (at least there are not examples
  highlighting applicability for the interaction of such  type).
 Thus, the problem of finding  exact solutions in a closed form for
   the DLV system (\ref{3-1})
modeling the interaction between preys and predators is still
unsolved. Probably, traveling fronts of the form (\ref{4-22}) are
the first example of such exact solutions.

Another problem of construction of exact solutions for the DLV type
systems arise when one examines such systems with time-delay in
order to take into account, for example, the age of species in the
population. Some examples are presented in the very recent paper
\cite{pol-sor-22}.

The authors  are grateful to Yurko Holovach (Lviv, NAS of Ukraine)
who brought our attention to the very old papers by Julius Hirniak
\cite{hirniak-1908,hirniak-1911}. R.Ch. is grateful to late Wilhelm
Fushchych who  encouraged  him   to study nonlinear systems of
reaction-diffusion PDEs using the  Lie symmetry method. Professor
Fushchych passed away 25 years ago and the authors would like to
dedicate this review to his memory.

\end{document}